\documentstyle[12pt,epsfig]{article}
\def\setboxz@h{\setbox\z@\hbox}% forgotten from two following files
\def\wdz@{\wd\z@}% forgotten from two following files
\expandafter\chardef\csname pre amssym.def at\endcsname=\the\catcode`\@
\catcode`\@=11
\def\undefine#1{\let#1\undefined}
\def\newsymbol#1#2#3#4#5{\let\next@\relax
 \ifnum#2=\@ne\let\next@\msafam@\else
 \ifnum#2=\tw@\let\next@\msbfam@\fi\fi
 \mathchardef#1="#3\next@#4#5}
\def\mathhexbox@#1#2#3{\relax
 \ifmmode\mathpalette{}{\m@th\mathchar"#1#2#3}%
 \else\leavevmode\hbox{$\m@th\mathchar"#1#2#3$}\fi}
\def\hexnumber@#1{\ifcase#1 0\or 1\or 2\or 3\or 4\or 5\or 6\or 7\or 8\or
 9\or A\or B\or C\or D\or E\or F\fi}
\font\tenmsa=msam10
\font\sevenmsa=msam7
\font\fivemsa=msam5
\newfam\msafam
\textfont\msafam=\tenmsa
\scriptfont\msafam=\sevenmsa
\scriptscriptfont\msafam=\fivemsa
\edef\msafam@{\hexnumber@\msafam}
\mathchardef\dabar@"0\msafam@39
\def\dashrightarrow{\mathrel{\dabar@\dabar@\mathchar"0\msafam@4B}}
\def\dashleftarrow{\mathrel{\mathchar"0\msafam@4C\dabar@\dabar@}}

\def\ulcorner{\delimiter"4\msafam@70\msafam@70 }
\def\urcorner{\delimiter"5\msafam@71\msafam@71 }
\def\llcorner{\delimiter"4\msafam@78\msafam@78 }
\def\lrcorner{\delimiter"5\msafam@79\msafam@79 }
\def\yen{{\mathhexbox@\msafam@55 }}
\def\checkmark{{\mathhexbox@\msafam@58 }}
\def\circledR{{\mathhexbox@\msafam@72 }}
\def\maltese{{\mathhexbox@\msafam@7A }}
\font\tenmsb=msbm10
\font\sevenmsb=msbm7
\font\fivemsb=msbm5
\newfam\msbfam
\textfont\msbfam=\tenmsb
\scriptfont\msbfam=\sevenmsb
\scriptscriptfont\msbfam=\fivemsb
\edef\msbfam@{\hexnumber@\msbfam}

\def\widehat#1{\setbox\z@\hbox{$\m@th#1$}%
 \ifdim\wd\z@>\tw@ em\mathaccent"0\msbfam@5B{#1}%
 \else\mathaccent"0362{#1}\fi}
\def\widetilde#1{\setbox\z@\hbox{$\m@th#1$}%
 \ifdim\wd\z@>\tw@ em\mathaccent"0\msbfam@5D{#1}%
 \else\mathaccent"0365{#1}\fi}
\font\teneufm=eufm10
\font\seveneufm=eufm7
\font\fiveeufm=eufm5
\newfam\eufmfam
\textfont\eufmfam=\teneufm
\scriptfont\eufmfam=\seveneufm
\scriptscriptfont\eufmfam=\fiveeufm

\catcode`\@=\csname pre amssym.def at\endcsname
\expandafter\ifx\csname pre amssym.tex at\endcsname\relax
                                         \else \endinput\fi
\expandafter\chardef\csname pre amssym.tex at\endcsname=\the\catcode`\@
\catcode`\@=11
\newsymbol\gtrsim 1326
\newsymbol\lesssim 132E
\catcode`\@=\csname pre amssym.tex at\endcsname
\newcommand{\cm}{\mbox{\rm ~cm}}
\newcommand{\meter}{\mbox{\rm ~m}}
\def\GeV{\hbox{$\;\hbox{\rm GeV}$}}

\newcommand{\picob}{\mbox{{\rm ~pb}}}
\newcommand{\nanos}{\mbox{{\rm ~ns}}}
\begin{document}
\begin{titlepage}
  \begin{flushleft}
{\tt DESY 97-24}\hfill {\tt ISSN 0418-9833} \\
{\tt February 13th 1997}
  \vspace*{2.5cm} 
\end{flushleft}
\vspace*{1.0cm}
\begin{center}
\begin{Large}
\boldmath
\bf{Observation of Events at Very High $Q^2$ 
    in $ep$ Collisions at HERA \\}
\unboldmath
 
\vspace*{1.5cm}
H1 Collaboration \\
\end{Large}
 
\vspace*{0.5cm}
 
\end{center}
 
\vspace*{1cm}
 
\begin{abstract}
\noindent
Measurements of $ep$ scattering with squared $4$--momentum transfer $Q^2$ 
up to 35000 \\
GeV$^2$ are compared with the expectation of the standard 
deep-inelastic model of lepton--nucleon scattering (DIS).
For $Q^2 > 15000\GeV^2$, $N_{obs} =12$ 
neutral current candidate events are observed where the expectation is 
$ N_{DIS} = 4.71 \pm 0.76 $ events. 
In the same $Q^2$ range, $N_{obs} =4$ charged current candidates are 
observed where the expectation is $ N_{DIS} = 1.77 \pm 0.87 $ events.
The probability ${\cal{P}}(N \geq N_{obs})$ that the DIS model signal 
$ N $ fluctuates to $ N \geq N_{obs} $ in a random set of experiments is 
$6 \times 10^{-3}$ for neutral current and $0.14$ for charged current.
The difference in the observed and expected number of Neutral Current
events is mostly due to events at large masses $M = \sqrt{x \, s}$
in which the positron is backscattered at large $y = Q^2/M^2$.
\vspace{1cm}
 
\centering{Submitted to Zeitschrift f\"ur Physik C}
 
\end{abstract}
\end{titlepage}
 
\vfill
\clearpage
 
%========= Standard H1 authors and institutes list =====================
%
\begin{flushleft}
  
%   H1AUTS  Author list by names, no. of authors  396
%           status: 15/01/97   13.30.23
 C.~Adloff$^{35}$,                %WUPP-ST                  Adloff              
 S.~Aid$^{13}$,                   %HAM2-LEFT    8/96        Aid                 
 M.~Anderson$^{23}$,              %MANC-ST  10/95           Anderson            
 V.~Andreev$^{26}$,               %LPI -PD                  Andreev             
 B.~Andrieu$^{29}$,               %ECPL-PD                  Andrieu             
 V.~Arkadov$^{36}$,               %ZEUT-ST    10/96         Arkadov             
 C.~Arndt$^{11}$,                 %DESY-ST   1/96           Arndt               
 I.~Ayyaz$^{30}$,                 %PARI-ST       5/96       Ayyaz               
 A.~Babaev$^{25}$,                %ITEP-PD                  Babaev              
 J.~B\"ahr$^{36}$
% ...

\end{flushleft}
\begin{flushleft} 
  {\it %     H1 Institutes as appearing on publications
% ...
$ ^2$ III. Physikalisches Institut der RWTH, Aachen, Germany$^ a$ \\
% ...
$ ^{11}$ DESY, Hamburg, Germany$^a$ \\
% ...
$ ^{13}$ II. Institut f\"ur Experimentalphysik, Universit\"at Hamburg,
                                     Hamburg, Germany$^ a$  \\
% ...
$ ^{23}$ Physics Department, University of Manchester,
                                          Manchester, UK$^ b$\\
% ...
$ ^{25}$ Institute for Theoretical and Experimental Physics,
                                                 Moscow, Russia \\
 $ ^{26}$ Lebedev Physical Institute, Moscow, Russia$^ f$ \\
% ...
$ ^{29}$ LPNHE, Ecole Polytechnique, IN2P3-CNRS,
                             Palaiseau, France \\
 $ ^{30}$ LPNHE, Universit\'{e}s Paris VI and VII, IN2P3-CNRS,
                              Paris, France \\
% ...
$ ^{35}$ Fachbereich Physik, Bergische Universit\"at Gesamthochschule
               Wuppertal, Wuppertal, Germany$^ a$ \\
 $ ^{36}$ DESY, Institut f\"ur Hochenergiephysik,
                              Zeuthen, Germany$^ a$\\
% ...
$ ^a$ Supported by the Bundesministerium f\"ur Bildung, Wissenschaft,
        Forschung und Technologie, FRG,
        under contract numbers 6AC17P, 6AC47P, 6DO57I, 6HH17P, 6HH27I,
        6HD17I, 6HD27I, 6KI17P, 6MP17I, and 6WT87P \\
 $ ^b$ Supported by the UK Particle Physics and Astronomy Research
       Council, and formerly by the UK Science and Engineering Research
       Council \\
% ...
$ ^f$ Supported by the Deutsche Forschungsgemeinschaft \\
% ...
 }
\end{flushleft}
 
%=======================================================================
\newpage
\section{Introduction}
\label{sec:intro}
 
The $ep$ collider HERA offers the unique possibility to probe the proton
at very small distances ($\simeq 10^{-16} \cm$) via $t$-channel 
exchange of highly virtual gauge 
bosons~\cite{H1ZEUSHQ2,H1Q2CC95,H1Q2CC96}, 
and to search in the $s$-channel for new particles which couple 
to lepton-parton pairs with masses up to the kinematic limit of 
$\sqrt{s} \simeq 300 \GeV$~\cite{H1ZEUSLQ}.

In this paper, measurements of neutral current (NC) and charged 
current (CC) deep-inelastic scattering (DIS) are considered using 
all available $e^+ p$ data collected by H1 at HERA from 1994 to 1996.
The positron beam energy $E^0_e$ was $27.5 \GeV$ and the proton 
beam energy $E_p$ was $820 \GeV$.
The total integrated luminosity amounts to $14.19\pm 0.32$ pb$^{-1}$,
an increase of a factor $\sim 5$ compared to the above cited earlier 
studies.

%=======================================================================
\section{The H1 Detector}
\label{sec:detect}
 
A detailed description of the H1 detector can be found
in~\cite{H1DETECT}.
Here we describe only the components relevant for the present analysis
in which the final state of the events involves either 
a positron\footnote{The analysis does not distinguish explicitly between 
                    $e^+$ and $e^-$.} 
with high transverse energy or a large amount of hadronic transverse 
energy flow.

The positron energy and angle are measured in a liquid argon (LAr) sampling 
calorimeter~\cite{H1LARCAL} covering the polar
angular\footnote{The $z$ axis is taken to be in the direction of the 
                 incident proton, and the origin of coordinates is 
                 the nominal $ep$ interaction point.}
range 4$^{\circ} \le \theta \le$ 153$^{\circ}$ and all azimuthal angles.
The granularity is optimized to provide fine and approximately uniform 
segmentation in laboratory pseudorapidity and azimuthal angle $\phi$.
It consists of a lead/argon electromagnetic section
followed by a stainless steel/argon hadronic section.
Electromagnetic shower energies are measured with a resolution of 
$\sigma(E)/E \simeq$ $12\%$/$\sqrt{E/\GeV} \oplus1\%$ and pion 
induced hadronic energies with 
$\sigma(E)/E \simeq$ $50\%$/$\sqrt{E/\GeV} \oplus2\%$ after 
software energy weighting.
These resolutions were measured in test beams with electron energies
up to 166 GeV~\cite{H1CALEPI,H1CALRES} and 
pion energies up to 205 GeV~\cite{H1CALRES}.
The absolute energy scales are known to 3\% and 4\% for
electromagnetic and hadronic energies respectively.
The angular resolution on the positron measured from the
electromagnetic shower in the calorimeter varies from $\sim 2$ mrad
below 30$^{\circ}$ to $\lesssim 5$ mrad at larger angles.
A lead/scintillating-fibre backward calorimeter~\cite{SPACAL} extends the 
coverage\footnote{The detectors in the backward region were upgraded 
                  in 1995 by the replacement of the lead/scintillator tile 
                  calorimeter~\cite{BEMC} and a proportional chamber.}
at larger angles (155$^{\circ} \le \theta \lesssim$ 178$^{\circ}$).
 
Located inside the calorimeters is the tracking system which is used here 
to determine the interaction vertex.
The main components of this system are central drift and proportional
chambers (25$^{\circ} \le \theta \le$ 155$^{\circ}$), a forward track
detector  (7$^{\circ} \le \theta \le$ 25$^{\circ}$) and a backward
drift chamber\footnotemark[3].
The tracking chambers and calorimeters are surrounded by a superconducting 
solenoid providing a uniform field of $1.15${\hbox{$\;\hbox{\rm T}$}}
parallel to the $z$ axis within the detector volume.
The instrumented iron return yoke surrounding this solenoid is used to 
measure leakage of hadronic showers and to recognize muons.
The luminosity is determined from the rate of the Bethe-Heitler
$e p \rightarrow e p \gamma$ bremsstrahlung measured in a luminosity 
monitor. This consists of a positron tagger ($e$-tagger) and a
photon tagger ($\gamma$-tagger), located $- 33 \meter$ and 
$- 103 \meter$ respectively from the interaction point. 

%=======================================================================     
\section{Event Selection and Kinematics}
\label{sec:select}

%====================================================
\subsection{Selection of {\boldmath $ep$} Collisions}

The analysis considers the accumulated data corresponding 
to H1 running conditions for which the central jet chambers (CJC), the LAr 
calorimeter and associated triggers, the backward calorimeter, the 
time-of-flight system and the luminosity system are fully operational.
The trigger requirement is either an electromagnetic cluster or a large
imbalance in transverse ``momentum'' $P^{trig}_{T,miss}$
measured with coarse trigger towers of the LAr calorimeter~\cite{H1LARCAL}. 
Every tower $k$ provides a fast energy sum ${\vec E}_k$ which contributes to 
the reconstruction of 
$P^{trig}_{T,miss} \equiv \sqrt{(\sum E_{x,k})^2+(\sum E_{y,k})^2}$.

Background not related to $e^+p$ collisions is first rejected by requiring 
for each event~: 
\begin{enumerate}
% 1
  \item a primary interaction vertex in the range
        $\mid z - \bar{z} \mid < 35 \cm$ 
        where $\bar{z}$ varies within $\pm 5 \cm$ around $z=0$
        depending on the HERA beam settings;        
% 2
  \item that the event time $t_0$ determined accurately 
        ($\sigma^{CJC}(t_0) \simeq 1.6 \nanos$) from the central jet 
        chambers coincides with the nominal time of the bunch crossings 
        of the beams;
% 3
  \item that it should survive a set of halo and cosmic muon 
        filters~\cite{H1Q2CC95}.
\end{enumerate}
Cut (1) mainly suppresses beam-wall and beam-residual gas
interactions. Cuts (2) and (3) eliminate any remaining
background produced by cosmic rays and by ``halo'' muons
associated with the proton beam.

%===================================================
\subsection{Neutral Current Selection and Kinematics}
\label{sec:ncdisevts}

The following criteria are designed to select NC DIS events in the high 
$Q^2$ domain.
The cuts, which rely only on calorimetric information, require :
\begin{enumerate}
% 1
  \item an isolated 
%  energy cluster identified as a 
        positron
%  ~\cite{H1CALEPI}
        with $E_{T,e} > 25 \GeV$ ($E_{T,e} = E_e \sin \theta_e$),
        found within the polar angular range 
        $10^{\circ} \le \theta_e \le  145^{\circ}$;
        the positron energy cluster should contain more than $98\%$ of 
        the LAr energy found within a pseudorapidity-azimuthal cone 
        of opening $\sqrt{ (\Delta \eta_e)^2 + (\Delta \phi_e)^2 } = 0.25$
        where $\eta_{e} = -\ln \tan \frac{\theta_e}{2}$; 
% 2
  \item a total transverse momentum balance 
        $P_{T,miss}/\sqrt{E_{T,e}} \leq 3 \sqrt{\GeV}$ where
        $P_{T,miss} \equiv  \sqrt{ \left(\sum E_{x,i} \right)^2 
                                +  \left(\sum E_{y,i} \right)^2 } $
        summed over all energy deposits $i$ in the
        calorimeters ($ E_{x,i} = E_i \sin \theta_i \cos \phi_i $
        and $ E_{y,i} = E_i \sin \theta_i \sin \phi_i $);
% 3
  \item a limited momentum loss in the direction of the incident positron,
        $ 43 \leq \sum \left(E - P_z\right) \leq 63 \GeV$,
        where $\sum \left(E - P_z\right) \equiv 
               \sum \left(E_i - E_{z,i}\right) $ with
        $ E_{z,i} = E_i \cos \theta_i$.
\end{enumerate}
The identification of positron induced showers
relies on detailed knowledge of the expected lateral and longitudinal 
shower properties~\cite{H1CALEPI}.
The efficiency for the detection of positrons exceeds $90\%$ everywhere 
within the acceptance cuts,   
the main losses being due to showers developing through the inactive 
material between calorimeter modules.
The cut (2) makes possible a very efficient NC DIS selection up to the 
highest $Q^2$ by taking into account the natural scale of the hadronic 
energy resolution.
The cut (3) retains $90\%$ of NC DIS events and exploits the fact that by 
energy-momentum conservation, the $\sum \left(E - P_z\right)$ 
distribution for DIS events is peaked at $2 E_e^0$. 
It rejects events where a very hard collinear $\gamma$ is emitted 
by the initial state positron.

To calculate the appropriate DIS Lorentz invariants $Q^2, y $ and 
$M=\sqrt{xs}$, two different sets of estimators are used.
Firstly, use is made of only the measurement of the scattered positron:
$$ M_e = \sqrt{ s x_e } = \sqrt{\frac{Q^2_e}{y_e}}, \;\;\;\;
     Q^2_e = \frac{E^2_{T,e}}{1-y_e}, \;\;\;\;
       y_e = 1 - \frac{E_e - E_e \cos \theta_e}{2E_e^0} \;\; . \;\;  $$
This method will henceforth be denoted the $e$-method.
Secondly use is made of the reconstructed angles $\theta_e$ and $\theta_h$ 
of the positron and hadronic final state.
Using $\alpha_e = \tan (\theta_e/2) = (E_e - E_{z,e}) / E_{T,e}$, and 
$\alpha_h = \tan (\theta_h/2) = \sum (E_i - E_{z,i}) /  
                  \sqrt{ (\sum E_{x,i})^2 + (\sum E_{y,i})^2 }$,
where the summations are over all energy deposits of the hadronic final state,
then~\cite{HOEGER}~:                        
$$ M_{2 \alpha} = 
      \sqrt{ \frac {s}{\alpha_e \alpha_h} \frac{E^0_e}{E_p} } , \;\;\;\;
   Q^2_{2 \alpha} = \frac{E^0_e}{E_p} 
                    \frac{s}{\alpha_e (\alpha_e + \alpha_h)}    , \;\;\;\;
   y_{2 \alpha} = \frac{\alpha_h}{\alpha_e + \alpha_h}  .$$      
This method will henceforth be denoted the $2\alpha$-method.
In the parton model, the variable $x$ is interpreted as the momentum 
fraction carried by a parton in the proton.
The centre of mass energy of the positron-parton collision is $M$.  
The variable $y$ is related to the polar scattering angle $\theta_e^*$
of the positron relative to the incident proton in the centre of mass
frame of the positron-parton collision according to 
$y=(1+\cos\theta^*_e)/2$.

In addition to the cuts (1) to (3) above, the analysis is restricted to 
the kinematic range $Q^2 > 2500 \GeV^2$ and $0.1 < y < 0.9$.
The resolution in $M_e$ degrades with decreasing $y_e$ 
($ \delta M_e / M_e \propto 1/ y_e$) and so the low 
$y$ domain is excluded.
Excluding the high $y$ values suppresses the photoproduction background 
(see section~\ref{sec:background}) and also avoids the region where 
QED radiative effects are largest for the $e$-method. 
For this kinematic range, the NC trigger efficiency exceeds $96\%$ and
is consistent with $100\%$ to within experimental error.

Following the NC selection described above, $471$ ($493$) DIS event 
candidates are accepted using the $e$-method ($2\alpha$-method) for
the calculation of the kinematic cuts.

%===================================================
\subsection{Charged Current Selection and Kinematics}

The selection of event candidates for the CC DIS sample requires:
\begin{enumerate}
% 1
  \item the total missing transverse momentum 
        $P_{T,miss} = P_{T,h} > 50 \GeV$;
% 2
  \item the total transverse energy
        $E_T=\sum \sqrt {E_{i,x}^2+E_{i,y}^2}$
        calculated from energy deposits in the calorimeters
        to match $P_{T,h}$ such that $(E_T - P_{T,h})/E_T < 0.5$.
\end{enumerate}
These cuts eliminate the photoproduction and NC DIS background. 

The DIS Lorentz invariants $Q^2$, $y$ and $M$ are calculated using 
the Jacquet-Blondel ansatz~\cite{HOEGER} by summing over 
all measured final state hadronic energy deposits using:
$$ M_h = \sqrt{\frac{Q^2_h}{y_h}}, \;\;\;\;
     Q^2_h= \frac{P^2_{T,h}}{1-y_h},\;\;\;\;
       y_h=\frac{\sum \left(E-P_z\right)}{2E_e^0}.\;\;\;\; $$
This method will henceforth be denoted the $h$-method.
The $h$-method can also be used for NC DIS.

In addition to the cuts (1) and (2) above, the analysis is restricted 
to the kinematic range $y_h < 0.9$.
The resolutions in both $M_h$ and $Q^2_h$ degrade with increasing $y$
(both $\delta M_h / M_h$ and $\delta Q^2_h / Q^2_h$ behave as
$\propto 1/(1-y_h)$ for $y_h \sim 1$)
and so the high $y_h$ domain is excluded.
The requirement of cut (1) restricts implicitly the kinematic range to
$Q^2_h >  2500 \GeV^2$. Throughout this range, the CC trigger efficiency 
is $\gtrsim 90 \%$.

The CC selection described above retains $31$ CC DIS event candidates.

%=======================================================================
\section{Standard Model of Deep-Inelastic Scattering}
\label{sec:dismc}

The calculation of an expectation for NC and CC DIS $ep$ scattering is 
performed using the parton model
in the approximation of single $\gamma/Z$ and $W$ boson exchange.
Detailed expressions for the NC and CC cross-sections which contain 
a description of the proton in terms of scale dependent structure 
functions can be found in~\cite{H1Q2CC96}.
For NC DIS, the contribution of the longitudinal structure function
$F_L$ is neglected.
The structure functions are expressed in terms of parton densities 
which are taken here from the Martin-Roberts-Stirling MRS~(H)~\cite{MRSH} 
parametrizations.
These were determined by a global fit of structure functions 
measurements in fixed target experiments~\cite{BCDMS,NMC,CCFR,EMCNA51}, 
inclusive lepton production and direct photon 
production~\cite{WA70,E605,CDFD0,UA2UA6}, and low $Q^2$ structure 
functions measurements at HERA~\cite{H1F293,ZEUSF293}.
The parton densities are evolved to the high $Q^2$ relevant for this 
analysis using the next-to-leading order DGLAP equations~\cite{DGLAP}
and divergences are regulated in the DIS renormalization 
scheme~\cite{DISCHEME}. 
Higher twist contributions are neglected.

For the comparison with data, a Monte Carlo simulation following such an 
approach is performed using the event generator DJANGO~\cite{DJANGO}.  
This generator includes the QED first order radiative corrections, 
and emission of real bremsstrahlung photons~\cite{HERACLES}. 
Moreover it offers an interface to two possible models for the 
emission of QCD radiation.
The LEPTO~\cite{LEPTO} model includes
the QCD matrix elements to first order in $\alpha_s$, complemented by
leading-log parton showers to model higher order radiation.
The ARIADNE~\cite{ARIADNE} generator makes use of the Color Dipole
Model~\cite{CDM} to simulate QCD radiation to all orders.
%PS/YS Is "all orders" OK
Both use string fragmentation~\cite{JETSET74} to
generate the hadronic final state. 

A complete Monte Carlo simulation of the H1 detector response 
is performed for NC and CC DIS processes. 
For NC DIS, a sample corresponding to $Q^2 > 1000 \GeV^2$
and which amounts to $\sim 25$ times the integrated luminosity of 
the data was produced using ARIADNE.
A similar sample was produced using LEPTO.
To reduce further the Monte Carlo statistical uncertainty at high $Q^2$,
a simulation corresponding to $\sim 100$ times the integrated 
luminosity of the data was performed using ARIADNE with the additional 
requirement $E_e > 50 \GeV$ on the energy of the scattered positron. 
The same event generator was used for the simulation of CC DIS events 
with $Q^2 > 100 \GeV^2$ to give a sample also $\sim 100$ times the 
integrated luminosity of the data.
For the comparison with  experimental data in the following sections,
the ARIADNE model for the hadronic final state is used unless
explicitly otherwise stated.

The following experimental errors are propagated as systematic errors on
the mean standard NC DIS expectation:
\begin{itemize}
  \item the uncertainty on the integrated luminosity ($\pm 2.3\%$);
  \item the uncertainty on the absolute calibration of the calorimeters 
        for electromagnetic energies ($\pm 3\%$);
  \item the uncertainty on the absolute calibration of the calorimeters 
        for hadronic showers ($\pm 4\%$); for NC DIS, this error only enters 
        via the constraints on $P_{T,miss}$ and on $\sum (E-P_z)$.
\end{itemize}
In addition, a $7\%$ ``theoretical'' uncertainty on the predicted NC DIS 
cross-section originates from contributions of:
\begin{itemize}
  \item the uncertainty of $5\%$ on parton density distributions extracted 
        from ``QCD fits''; this is partly due to the experimental errors on
        the input data (in particular in the structure function measurements
        in the high $x$ range from the BCDMS experiment~\cite{BCDMS}) and
        is partly linked to the assumptions for the shapes of the parton 
        distributions at the $Q^2_0$ at which the perturbative QCD 
        evolution is started; 
        this uncertainty is compatible with what can be inferred from a
        comparison of the cross-section predictions, calculated with 
        HECTOR~\cite{HECTOR}, using the recent sets of next-to-leading 
        order parametrizations of the MRS~\cite{MRSR12}, 
        CTEQ~\cite{CTEQ4MHJ}, and GRV~\cite{GRV94HO} groups regulated 
        in the $\overline{MS}$ scheme~\cite{MSBAR};
%YS> The GRV parametrizations is based on the dynamical parton 
%YS> model~\cite{DYNPARTON}. 
  \item the value of the strong coupling constant $\alpha_s$, which leads 
        to an uncertainty of $4\%$; this was inferred by comparing the 
        CTEQ~(A1) to (A5)~\cite{CTEQA1A5} parametrizations with
        $\alpha_s(M_Z)$ ranging from $0.110$ to $0.122$; 
        a similar value can be inferred by a
        comparison of the MRS~(R1) and (R2)~\cite{MRSR12} parametrizations; 
  \item the higher order QED corrections imply a $2\%$ uncertainty
        in the $y$ range considered in this analysis;
        this was estimated using HECTOR~\cite{HECTOR} which
        makes possible two approaches:
        QED radiative corrections can be calculated~\cite{HELIOS}  
        in the leading logarithmic approximation at 
        ${\cal {O}} (\alpha_{em})$ including ${\cal {O}} (\alpha_{em}^2)$ 
        corrections in the next-to-leading logarithmic approximation 
        as well as soft photon exponentiation; alternatively~\cite{TERAD},
        complete QED and electroweak corrections are calculated at
        ${\cal {O}} (\alpha_{em})$.
\end{itemize}
All analyses described in the following sections have been 
repeated with an independent shift of the central values by $\pm 1$ 
standard deviation of each of the experimental and theoretical sources of 
errors. The overall systematic error of the standard DIS model prediction 
is determined as the quadratic sum of the resulting errors 
and of the statistical error on the Monte Carlo simulation.

%=======================================================================
\section{Background Sources}
\label{sec:background}

The contributions from background processes which could give 
rise to events with true or misidentified isolated positrons at 
high $E_T$ or to events with a large $P_{T,miss}$ were evaluated 
using the full simulation and reconstruction program chain. 
The list of processes is given in Table~\ref{tab:background}.

The production of real electroweak vector bosons $Z^0$ and $W^{\pm}$
was modelled using the EPVEC event generator~\cite{H1EPVEC}.
This includes processes where the partonic structure of a photon is
resolved.
It should be noted that inelastic single $Z^0$ and $W^{\pm}$ production is 
not contained in the standard DIS model described in section~\ref{sec:dismc} 
and thus it is treated here as a background source. 
Especially  contributions from radiation of a $Z^0$ or $W^{\pm}$
from a quark can give rise to forward positrons and thus 
can mimic NC DIS events at very high $Q^2$.

Direct and resolved photoproduction processes were modelled using the
PYTHIA generator~\cite{PYTHIA}.
It is based on leading order QCD matrix elements and includes initial 
and final state parton showers calculated in the leading logarithm 
approximation, and string fragmentation~\cite{JETSET74}.
To enhance the generated yield of events in the hard scattering region,
the transverse momenta ${\hat p}_T$ in the hard sub-processes 
are required to exceed $10 \GeV$ for light quark flavour 
production, $3 \GeV$ for prompt photon processes, and $8 \GeV$ for heavy 
quark flavour production. 
These cuts do not affect the production of jets with 
$P_T > 25 \GeV$ (or of jets containing hard fragments with 
$ \sum_{e,\gamma} P_T > 20 \GeV $), or the production of prompt photons 
with $P_T > 10 \GeV$. 
The renormalization and factorization scales are both set to $P_T^2$.
The GRV~(G) leading order parton densities in the photon are used~\cite{GRVG}.
%
% ---------------------- TABLE 1: Generators ---------------------------
%
\begin{table*}[t]
  \renewcommand{\doublerulesep}{0.4pt}
  \renewcommand{\arraystretch}{1.2}
 \begin{center}
 \vspace{-1.3cm}
 \begin{tabular}{||c|c|c|c||}
 \hline \hline
 Partonic Process & Generator &  Simulated      &  Upper Limits (95\% CL)  \\
    (example)     & Model     &  Luminosity     &  on Events per     \\
                  & [refs.]   & ($\picob^{-1}$) &  14.19 pb$^{-1}$   \\ \hline
   \multicolumn{4}{||c||}{Single W and Z Boson Production}         \\ \hline
   \begin{tabular}{cclll}
  $e + q$     & $\longrightarrow$
              & $W$  & + & $ e + X $      \\
              & & ${\hookrightarrow}$  & &$ e + \nu $ \\
              & & ${\hookrightarrow}$  & &jet + jet   \\
   \end{tabular}
                     & \cite{H1EPVEC}
                              & \begin{tabular}{c}   \\
                                   $8600$ \\             
                                   $1400$ \\
                                \end{tabular}
                                       & \begin{tabular}{c}   \\
                                            $0.005$           \\
                                            $0.16$            \\
                                         \end{tabular}        \\
\hline
  \begin{tabular}{cclll}
  $e + q$     & $\longrightarrow$
              & $Z$  & + & $ e + X $      \\
              & & ${\hookrightarrow}$  & &$ e^+ + e^- $ \\
              & & ${\hookrightarrow}$  & &$\tau^+ + \tau^- $ \\
              & & ${\hookrightarrow}$  & & jet + jet         \\
  \end{tabular}
                      & \cite{H1EPVEC}
                             & \begin{tabular}{c}   \\
                                  $42000$ \\    
                                  $59000$ \\    
                                   $3600$ \\
                                \end{tabular}
                                       & \begin{tabular}{c}  \\
                                            $0.003$          \\
                                            $0.001$          \\
                                            $0.18$           \\
                                         \end{tabular}       \\
\hline
   \multicolumn{4}{||c||}{High $P_T$ jets photoproduction
                          (direct and resolved) }      \\ \hline
   $\gamma + q \rightarrow  q + g$,  \,
   $\gamma + g \rightarrow  q+ {\bar q}'$   
                               &          &           & \\
   \hfill (hard jet)  & \cite{PYTHIA}
                                   & 500  & $ 0.38 $   \\
   \hfill (hard fragmentation)
                      &            & 500  & $ 0.31 $   \\
\hline
   \multicolumn{4}{||c||}{Heavy Flavour Production 
                          (direct and resolved)}             \\ \hline
   $ \gamma + g \longrightarrow c + \bar{c} $
                      & \cite{PYTHIA}
                                 & 600    & $ 0.07 $   \\
   $ \gamma + g \longrightarrow b + \bar{b} $
                      & 
                                 & 600    & $ 0.07 $   \\
\hline
  \multicolumn{4}{||c||}{Prompt Photon Production
                         (direct and resolved)}              \\ \hline
$ \gamma + q \rightarrow q + \gamma$
                                  & \cite{PYTHIA}
                                       & 500   & $ 0.09 $       \\ \hline
 \multicolumn{4}{||c||}{ Two-photon processes}                  \\ \hline
 $ \gamma + \gamma \longrightarrow e^+ + e^- $
                      & \cite{LPAIR}
                                 & 4500  &  $  0.04 $   \\     
 $ \gamma + \gamma \longrightarrow  q + {\bar q} $
                      &          & 17000 &  $ 0.002 $   \\
\hline \hline
  \end{tabular}
         \caption {\small \label{tab:background}
                   Upper limits ($95\%$ CL) on the expected number of 
                   events from background processes which survive the event 
                   selection for NC DIS candidates and the full set of
                   background rejection cuts;
                   for the photoproduction of high $P_T$ jets
                   involving light quark flavours, two alternative selections 
                   are made which require either a hard jet or jets 
                   containing hard leptons, photons or $\pi^0$'s. }
 \end{center}

\vspace{-0.5cm}
\end{table*}
%
% ------------------------------------------------------------------------

Contributions from two-photon ($\gamma\gamma$) processes, where one
$\gamma$ originates from the proton, were also considered. 
Electron-positron pair production $e^+ p \rightarrow e^+ + e^+ e^- + X$, 
where at least one of the leptons has high $P_T$, was simulated
using the LPAIR generator~\cite{LPAIR}. LPAIR includes the 
relevant Bethe-Heitler diagrams to lowest order in 
$\alpha_{em}$ (i.e. order $\alpha_{em}^4$) but 
does not include the photon bremsstrahlung diagrams where the photon 
converts into a lepton pair or any diagrams in which real or 
virtual photons are replaced by $Z^0$ bosons.
The contribution from $q\bar{q}$ production in $\gamma\gamma$ collisions 
was simulated using PYTHIA~\cite{PYTHIA} by folding in appropriate $\gamma$
fluxes~\cite{GFLUXES},
and it did not include diagrams involving massive electroweak bosons.
The diagrams for both final states ($e^+e^-$ and $q \bar{q}$) which 
involve the formation of a $Z^0$ vector boson are included in
the calculations of electroweak vector boson production using the EPVEC
generator.

When the high $Q^2$ DIS selection criteria which are described in 
section~\ref{sec:select} are applied to the above background sources,
the total remaining background contamination is found to be below 
$1 \%$ (95\% CL). 
Nevertheless, in order to ensure that the background contamination is 
negligible everywhere in the kinematic plane compared with the 
uncertainty on the standard DIS expectation, additional cuts are imposed.
Each cut is specifically designed to eliminate a given background source.
% These cuts are specifically designed to eliminate a given background source.

Background contamination in the NC DIS selection is suppressed by the 
following:  
%%%<
\begin{enumerate}
% 1
  \item processes leading to multi-lepton final states are eliminated   
        by requiring at least one reconstructed jet with 
        $E_{T,jet} > 7 \GeV$ in the polar angular range 
        $10^{\circ} \le \theta_{jet} \le  145^{\circ}$;
        the jet is specified using a cone algorithm with a radius 
        $\sqrt{\Delta \eta_{jet}^2+ \Delta\phi_{jet}^2} = 1$;
        at least $1\%$ of the jet energy should be deposited in the 
        hadronic section of the LAr calorimeter; 
% 2
  \item QED Compton events are suppressed by rejecting events in 
        which a second positron (or photon) cluster with 
        $E_{T,e(2)} / E_{T,e} > 0.9$
        is found back-to-back in azimuth within 
        $\Delta \phi_{e,2} > 160^{\circ}$;
% 3
  \item background from single $Z^0$ boson production followed by a 
        $Z^0 \rightarrow e^+ e^-$ decay is suppressed 
        by rejecting events in which the invariant mass of
        two electron/positron clusters lies within $5 \GeV$ of the
        $Z^0$ boson mass;
% 4
  \item background from single $Z^0$ boson production followed by a 
        $Z^0 \rightarrow \tau^+ \tau^-$ decay is eliminated by the
        requirement 
        $\sum \left(E - P_z\right) > 0.2 \times P_{T,miss} + 43 \GeV$, 
        which rejects events having both abnormally high 
        $P_{T,miss}$ and abnormally low $\sum \left(E - P_z\right)$;
% 5
  \item background from isolated prompt photons in photoproduction 
        processes is eliminated by requiring that there be at least
        one charged track within the positron isolation cone;
% 6
  \item background from photoproduction and low $Q^2$  NC DIS 
        are removed by the requirement that there be less than $5 \GeV$ 
        in the backward calorimeters;
        moreover there should be no energy $E_{e-tag}$ in the $e$-tagger, 
        unless $\mid  E_{e-tag} + E_{\gamma-tag} - E^0_e \mid < 5 \GeV$,
        where $E_{\gamma-tag}$ is the energy in the $\gamma$-tagger,
        in order to save NC DIS events which coincide randomly with 
        Bethe-Heitler bremsstrahlung;
% 7
  \item background from $e^+e^-$ pair production in $\gamma$-$\gamma$ 
        collisions is strongly suppressed by requiring that there be in 
        addition to the ``scattered'' positron at most one 
        electron/positron candidate with $E_{T} > 4 \GeV$.
\end{enumerate}
The cuts (2), (3), (4), (5) and (7) each imply individually an 
efficiency loss for NC DIS processes below $1\%$ while the corresponding
losses for cuts (1) and (6) are below $2\%$.
Globally, the complete set of background rejection cuts leads to a
$5\%$ efficiency loss for NC DIS.
It should be noted that the background originating from single $Z^0$ 
or $W^{\pm}$ production followed by hadronic decay of the boson, and leading to
multi-jet final states, could be considerably reduced by requiring 
that at most one jet carries $E_{T,jet} > 40\% E_{T,e}$. 
This requirement has not been imposed in this analysis in order to stay as 
close as possible to an inclusive DIS measurement.
Applying all selection cuts and background rejection cuts, the remaining 
background contamination is below $0.1 \%$ (95\% CL) for the NC DIS sample.
Hence, the background is negligible when compared to the estimated errors
on the standard NC DIS expectation. 
This statement is also valid in a restricted kinematic range of
$Q^2 > 10000 \GeV^2$ where less than $0.1$ (95\% CL) background events are 
expected.
In most cases, the estimation of upper limits for the various background
sources in Table~\ref{tab:background} are limited by available Monte Carlo 
statistics.

To deal with specific background sources to CC DIS, it is  
required that there should be no isolated positron with 
$ E_{T,e} > 10 \GeV $ within $10^{\circ} \le \theta_e \le  145^{\circ}$. 
This rejects in particular background from single $W$ boson production 
followed by a $W \rightarrow e \nu$ decay. 
This causes negligible efficiency losses for the CC DIS selection,
while the remaining contamination from single $W$ production is below
$0.02 \%$ (95\% CL) within the analysis cuts.

The purity of the event selection is cross-checked by two methods:
\begin{enumerate}
\item for the NC selection, the positron isolation criteria were relaxed and 
      the positron cluster to charged track link was not applied; 
      no evidence for background was found;
\item all events from the NC and CC selection were scanned visually; 
      in the selected NC sample, no contamination or overlap with
      cosmic ray showers or beam halo events was seen; 
      there was no evidence for wrongly identified positron clusters;
      the selected CC sample showed no sign of contamination due to cosmic 
      ray showers; no muon candidate was found in any of the selected events.
\end{enumerate}

%=========================================================================
\section{Analysis}
\label{sec:analysis}

\subsection{General Properties of NC and CC Events}
\label{sec:general}

The measured data are compared with the expectation from the standard DIS 
model. Here and in all that follows, the expected yields of events 
are calculated using Monte Carlo simulation of the H1 detector response, 
including all detector and selection efficiencies, and the full H1 event 
reconstruction program. 
All comparisons with measurement are made with a normalization 
specified by the total cross-section within acceptance and the 
measured integrated luminosity. 

We find in the experimental data 443 (460) NC DIS candidates satisfying the
selection criteria and $Q^2_e>2500\GeV^2$ ($Q^2_{2\alpha}>2500\GeV^2$), which
is in good agreement with the expectation of $427 \pm 38$ ($442 \pm 40$)
from NC DIS processes.
Within the kinematic range in $\theta_e$, $E_{T,e}$, $y_e$ and $Q^2_e$
considered, the ratio of this expected number of events to the number
generated by the Monte Carlo simulation is about $0.8$ due to
experimental acceptance and event selection.
The sources of errors for the NC DIS expectation are discussed above.
For the CC DIS candidates, 31 events are found satisfying the above 
% --------------- FIGURE 1: control plots -----------------------------
 \begin{figure}[h]
 \vspace{-1.0cm}
 
   \hspace{2.0cm} \epsfxsize=0.7\textwidth 
   \epsffile{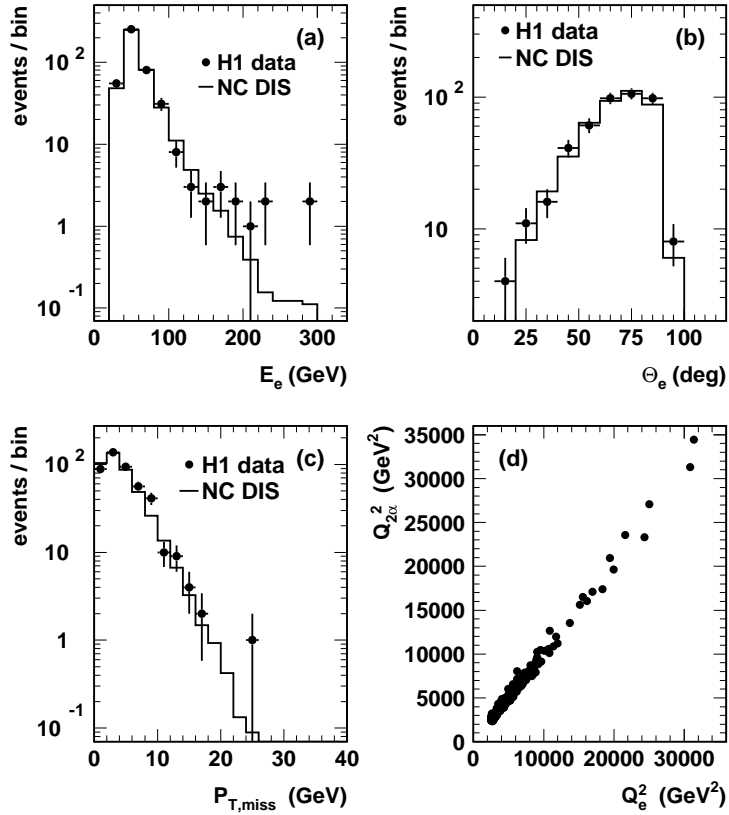}
   \caption[]{ \label{fig:check}
      {\small For NC DIS candidates, the distributions of (a) the measured 
              positron energy $E_e$, (b) the polar angle $\theta_e$, 
              (c) the missing transverse momentum $P_{T,miss}$,
              and (d) the correlation of $Q^2_{2\alpha}$ and $Q^2_e$, 
              (symbols); 
              in (a), (b) and (c)  the expectations from standard NC DIS
              assuming the integrated luminosity of the measurement are 
              shown as superimposed histograms. }}
  \end{figure}
% -------------------------------------------------------------------- 
requirements, which agrees well with the expectation of 
$34.2 \pm 5.8$ from CC DIS processes.
Here the corresponding ratio of accepted to generated events is again
about $0.8$ in the $P_{T,miss}$ and $y_h$ range considered.

Fig.~\ref{fig:check} shows, for NC DIS candidates, the distributions of energy 
and polar angle of the scattered positron, the missing transverse momentum,
and the correlation of the 4-momentum transfer squared $Q^2$ calculated with 
the $e$-method ($Q^2_e$) and the $2\alpha$-method ($Q^2_{2\alpha}$).
Superimposed as histograms are the expectations from the NC DIS 
simulation. 
The distributions are well described by the simulation except for the  
largest values of the measured positron energy $E_e$.
This will be quantified below in terms of $Q^2$ to which $E_e$ is
closely related at large values.
A good correlation is seen between the measurement of $Q^2$ by the 
electron and the $2\alpha$-method.

\begin{figure}[h]
 \vspace{-1.0cm}

%  \hspace{2.0cm} \epsfxsize=0.7\textwidth \epsffile{fig.v03.massmeth.eps}
   \hspace{2.0cm} \epsfxsize=0.7\textwidth \epsffile{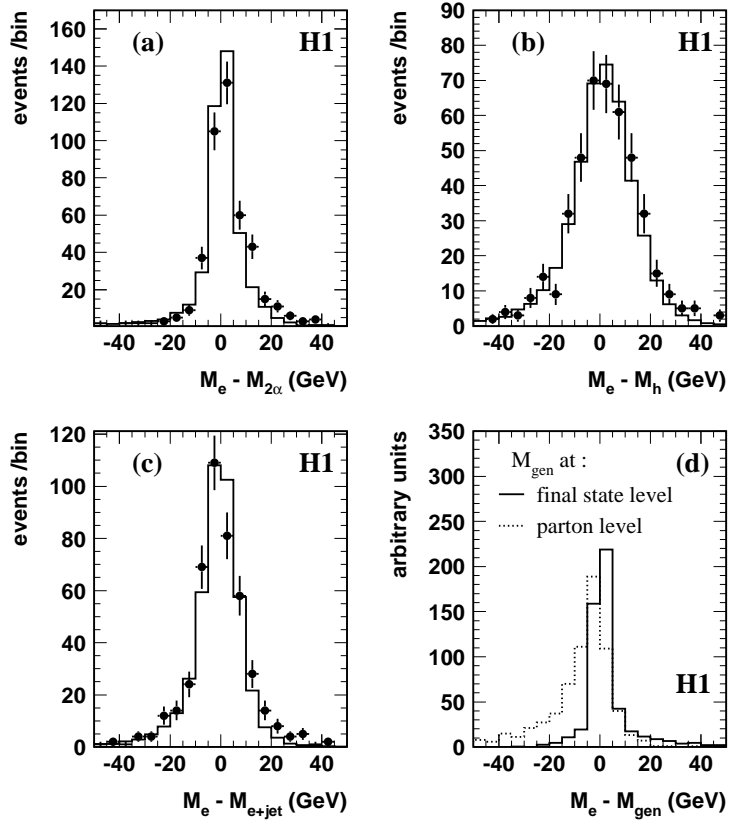}

   \caption[]{ \label{fig:massmeth}
      {\small Distributions for the NC DIS sample of the differences between 
              the mass ($M=\sqrt{xs}$) calculated using the $e$-method ($M_e$) 
              and (a) using the $2\alpha$-method ($M_{2\alpha}$), (b) using 
              the $h$-method ($M_h$), and (c) using the invariant mass of 
              the positron-jet system ($M_{e+jet}$); superimposed on
              the data points ($\bullet$ symbols) are histograms of the 
              expectation from the Monte Carlo simulation; 
              (d) difference between $M_e$ and the generated mass $M_{gen}$
              when the latter is calculated either from the final state 
              positron using simulated NC DIS events (histogram), or taken 
              from the invariant mass of the positron-quark sub-system in 
              a specific $eq$ resonance model (dotted histogram). }}    
\end{figure}

Fig.~\ref{fig:massmeth} shows the measured and expected differences 
between various mass reconstruction methods. 
%--------------- FIGURE 2: Mass Methods  ---------------------------
% --------------------------------------------------------------------
The $e$-method, which provides good resolution for all kinematic 
quantities at $y_e>0.1$, is taken here as a reference.
Good agreement is found between the mass measurement $M_e$ based on 
the $e$-method and $M_{2 \alpha}$ based on the $2\alpha$-method 
(Fig.~\ref{fig:massmeth}a). The difference $M_e - M_{2\alpha}$ is found
to be described acceptably by the NC DIS simulation.
The measurement $M_e$ is compared to $M_h$ calculated from the hadronic
energy flow in Fig.~\ref{fig:massmeth}b. Also here good agreement is
found between the two methods and the shape of the $M_e - M_h$ 
distribution is well described by the simulation.
This confirms the validity of the $h$-method which is the only one 
available to reconstruct the kinematics of CC DIS events.
A comparison of $M_e$ with the invariant mass $M_{e+jet}$ calculated from 
the energies and angles of the final state positron and highest $E_T$ 
jet is shown in Fig.~\ref{fig:massmeth}c.
Good agreement is found between the two mass methods which reflects the 
fact that the measurement is dominated by hard 
$e + parton \rightarrow e + parton$ 
scattering leading most often to a hadronic final state with a single 
high transverse energy jet. 
Also the difference $M_e - M_{e+jet}$ is well described by the simulation.

Fig.~\ref{fig:massmeth}d shows the expected experimental contribution to 
the resolution on $M_e$ obtained by comparing for the NC DIS simulation
the reconstructed and generated positron momentum 4-vectors.
The mean value of the distribution of the reconstructed 
$M_e$ is within $\pm 1\%$ of the true value of the Lorentz 
invariant $M = \sqrt{x s}$ calculated from the generated final state 
positron for the range of $y$ considered.
With respect to this generated positron, values of $M_e$ are
measured with an RMS resolution of $\sigma_{RMS} \simeq 7 \GeV$ and a 
Gaussian peak resolution of $\sigma_{Gauss} \simeq 3.0 \GeV$.
Also shown as a dotted histogram is the expected resolution of the
$e$--method for the invariant mass of the positron--parton system in a 
specific model (see section~\ref{sec:anomalous}).
It should be recalled that the $M_e$ resolution severely degrades at 
low $y_e$ and the events in Fig.~\ref{fig:massmeth} are restricted 
to $y_e > 0.1$.
A shift equal to the experimental systematic error on the calorimeter energy 
scale for positrons of $3\%$ would lead to  a $y$ dependent shift for the 
mass $M_e$ of $\delta M_e / M_e = 0.03/(2y_e).$

\begin{figure}[t]

   \hspace{2.0cm} \epsfxsize=0.7\textwidth \epsffile{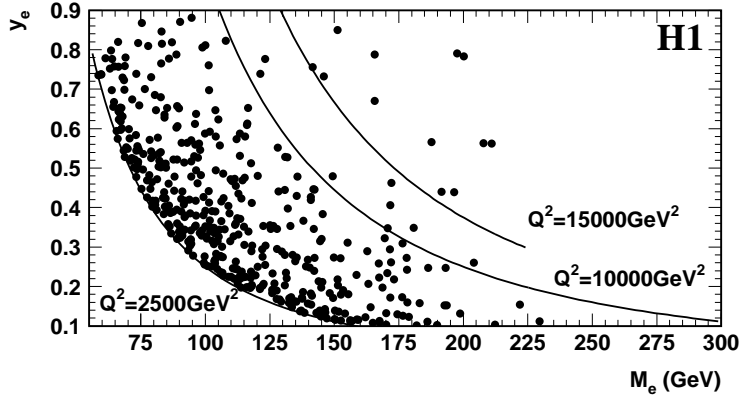}

   \caption[]{ \label{fig:scatter}
       {\small  Selected NC DIS candidate events in the 
                $M_e$ - $y_e$ plane;
                three contours of fixed $Q^2$ are shown. }}
 \end{figure}

Fig.~\ref{fig:scatter} shows the two dimensional distribution of $y_e$ 
against $M_e$.
% --------------- FIGURE 3: scatter plot -----------------------------
% --------------------------------------------------------------------
The cut-off in event density for low $y_e$ or $M_e$ arises because of the 
requirement $Q^2_e > 2500 \GeV^2$.
Also shown are the $Q^2_e = 10000 \GeV^2$ and $Q^2_e = 15000 \GeV^2$ 
contours.

\begin{figure}[htb]
\vspace{-0.5cm}

 \hspace{2.0cm} \epsfxsize=0.7\textwidth \epsffile{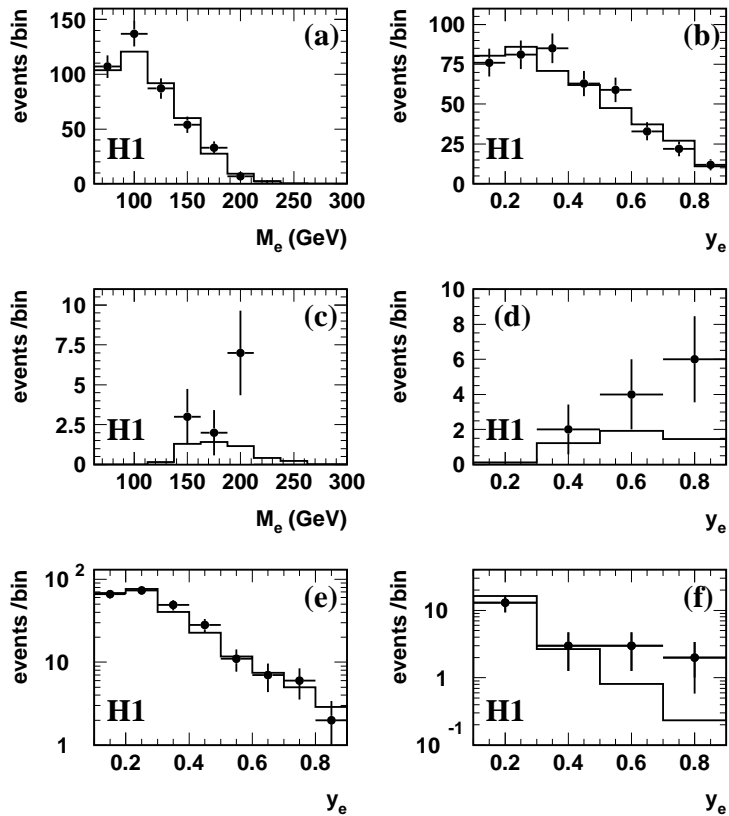}
 \caption[]{ \label{fig:mandy}
    { \small Distributions of $M_e$ and $y_e$ for selected NC DIS candidate 
             events, (a) and (b) for $2500 \GeV^2 < Q^2_e < 15000 \GeV^2$, 
             (c) and (d) for $Q^2_e > 15000 \GeV^2$; 
             distribution of $y_e$ (e) for $ 100 < M_e < 180 \GeV $ and 
             (f) for $ M_e > 180 \GeV $; superimposed on the data 
             points ($\bullet$ symbols) are histograms of the expectation 
             of standard NC DIS. }} 
\end{figure}

Fig.~\ref{fig:mandy}a and~\ref{fig:mandy}b show the projected $M_e$ and 
$y_e$ distributions of the selected events at ``low'' $Q^2$
($2500 \GeV^2 < Q^2_e < 15000 \GeV^2$) 
and Fig.~\ref{fig:mandy}c and~\ref{fig:mandy}d at ``high'' $Q^2$ 
($Q^2_e > 15000 \GeV^2$). 
The distributions of the data are well reproduced by standard DIS 
predictions in the low $Q^2$ range.
%
%---------------- FIGURE 4: M and y plots, for Q2 < 10000 and Q2 > 10000
%
%----------------------------------------------------------------------
At high $Q^2$ the data exceed the NC DIS expectation, 
especially at $M_e \sim 200 \GeV$.
Fig.~\ref{fig:mandy}e and~\ref{fig:mandy}f show the $y_e$ distributions
for $M_e$ values below and above $180 \GeV$. 
At high $M_e$, a difference between experiment and expectation 
is apparent at large $y_e$.

%========================================
\subsection{{\boldmath $Q^2$} Dependence}

%-------------------------------------------------
\subsubsection{Neutral Current Sample}
\label{sec:NCsample}

\begin{figure}[htb]

% \hspace{2.0cm} \epsfxsize=0.7\textwidth \epsffile{fig.v03.q2y01.eps}
  \hspace{2.0cm} \epsfxsize=0.7\textwidth \epsffile{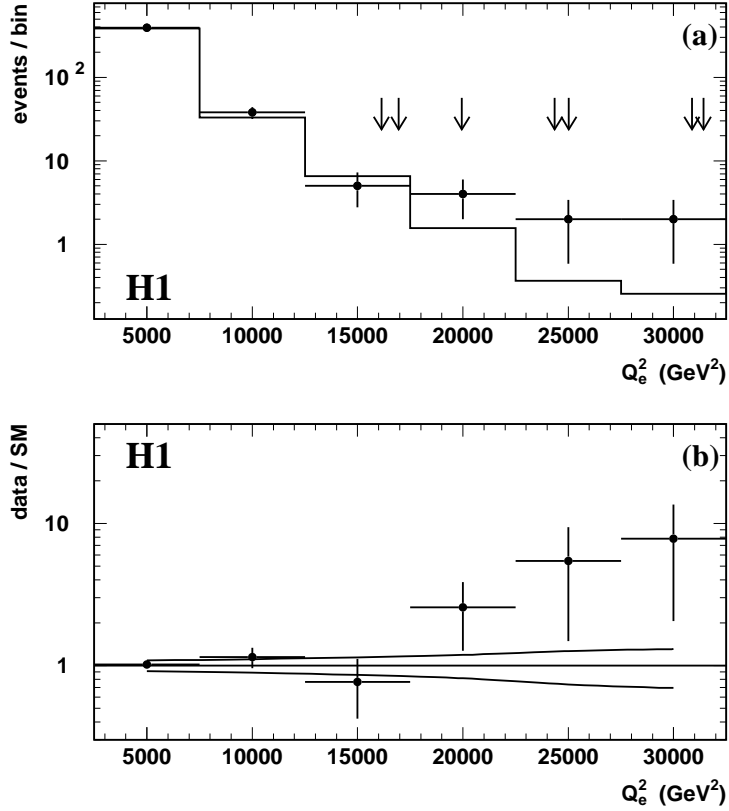}

   \caption[]{ \label{fig:q2plot}
      { \small (a) $Q^2_e$ distribution of the selected NC DIS candidate 
                events for the data ($\bullet$ symbols) and for standard 
                NC DIS expectation (histogram);
                the arrows indicate the $Q^2_e$ values for data 
                entries with $Q^2_e > 15000 \GeV^2$ and $M_e > 180 \GeV$;
                (b) ratio of the observed and expected number of
                events as a function of $Q^2_e$; 
                the lines above and below unity specify the 
                $\pm 1\sigma$ levels determined using the combination 
                of statistical and systematic errors of 
                the NC DIS expectation. }}

\end{figure}

Fig.~\ref{fig:q2plot} shows for the NC selection the
measured $Q^2_e$ distribution in comparison with the expectation 
from standard NC DIS processes.
Also shown is the ratio of the $Q^2_e$ distribution to the NC DIS 
expectation. 
Very similar results are obtained for this ratio for minimum values of 
$y_{min}$ in the range $0.1$ to $0.5$.
% --------------- FIGURE 5: Q2 plot y > 0.1 NC selection  -----------------
%
%------------------------------------------------------------------
The errors resulting from the convolution of the statistical error of the 
Monte Carlo sample and the systematic errors are correlated for
different $Q^2_e$ bins and are indicated in Fig.~\ref{fig:q2plot}b.
They are shown in this figure (and in subsequent figures~\ref{fig:ccq2plot} 
and~\ref{fig:massy0204}) as lines above and below unity joining the 
$\pm 1\sigma$ errors evaluated at the center of each bin.
These errors are dominated by the uncertainty in the electromagnetic 
energy scale of the calorimeter and vary between $8.5\%$ at low 
$Q^2_e$ and $30\%$ at the highest values of $Q^2_e$.
The NC DIS expectation agrees well with the data for 
$Q^2_e \lesssim 15000 \GeV^2$ while at larger $Q^2_e$
the number of events is in excess of the NC DIS expectation.

To quantify this difference, the numbers of observed and 
expected events with $Q^2_e$ above various $Q^2_{min}$ values are 
given in Table~\ref{tab:tabq2}.
Also given in Table~\ref{tab:tabq2} are the Poisson probabilities 
${\cal{P}}( N \geq N_{obs} )$ that in a random set of experiments 
the number of NC DIS events $N$ fluctuates to values equal to or 
larger than the observed number of events $N_{obs}$. 
% ---------------------- TABLE 2: Q^2 table NC -------------------------
\begin{table*}[htb]
  \renewcommand{\doublerulesep}{0.4pt}
  \renewcommand{\arraystretch}{1.2}
 \begin{center}
 \begin{tabular}{||l|c|c|c|c|c|c||}
 \hline \hline
$Q^2_{min} (\GeV^2)$  & 2500   & 5000   & 10000  & 15000  & 20000 & 30000 \\
                                                            \hline \hline
$N_{obs}$             & 443    & 122    & 20     & 12     & 5  & 2   \\
                                                            \hline
$N_{DIS}$          & 426.7 & 116.2 & 18.3  & 4.71   & 1.32  & 0.23 \\
    & $\pm 38.4$ & $\pm 11.6$ & $\pm 2.4$  & $\pm 0.76$ 
                                           & $ \pm 0.27$ & $ \pm 0.05$ \\
                                                            \hline
${\cal{P}}(N \geq N_{obs})$
                      & 0.35   & 0.35   & 0.39   & $6 \times 10^{-3}$  
                      & $1.4 \times 10^{-2}$   &  $2.3 \times 10^{-2}$  \\
 \hline \hline
  \end{tabular}
         \caption {\small \label{tab:tabq2}
                   Number of observed ($N_{obs}$) and expected NC DIS
                   ($N_{DIS}$) events with $4$-momentum transfer squared 
                   $Q^2_e$ above given thresholds ($Q^2_{min}$);
                   ${\cal{P}}(N \geq N_{obs})$ is the probability that the
                   number of NC DIS events fluctuates to values
                   equal to or larger than $N_{obs}$ in a random set of
                   experiments.}

   \end{center}
\end{table*}
% ------------------------------------------------------------------------
The systematic error $\delta b$ on the mean number of expected events $b$                                       
is taken into account by using the convolution~:
$$ {\cal{P}}(N \geq N_{obs} ) = \int_{0}^{+\infty} dx G(x;b,\delta b)
                                \sum_{k=N_{obs}}^{\infty} p(k;x) $$
where the following notations have been introduced~:
\begin{itemize}                                                                 
 \item $p(k;x)$ is the Poisson probability to observe $k$ events
       when the number of expected events is $x$, i.e.
       $p(k;x) = e^{-x} x^k / k! $;
 \item $G(x;b,\delta b)$ is the probability density function for the 
       NC DIS expectation $x$, namely a Gaussian of mean 
       value $b$ and width $ \delta b$.                                         
\end{itemize}                                   
The resulting probabilities are about $1\%$ at the 
largest $Q^2_{min}$ values.
For $Q^2 > 15000 \GeV^2$, the number of observed events 
is $N_{obs}=12$ for an expectation of $4.71 \pm 0.76$ 
corresponding to a probability ${\cal P}(N \geq N_{obs}) $ of 
$6 \times 10^{-3}$. 

%------------------------------------------------------
\subsubsection{Charged Current Sample}
\label{sec:CCsample}

\begin{figure}[htb]
\vspace{-0.5cm}

   \hspace{2.0cm} \epsfxsize=0.7\textwidth \epsffile{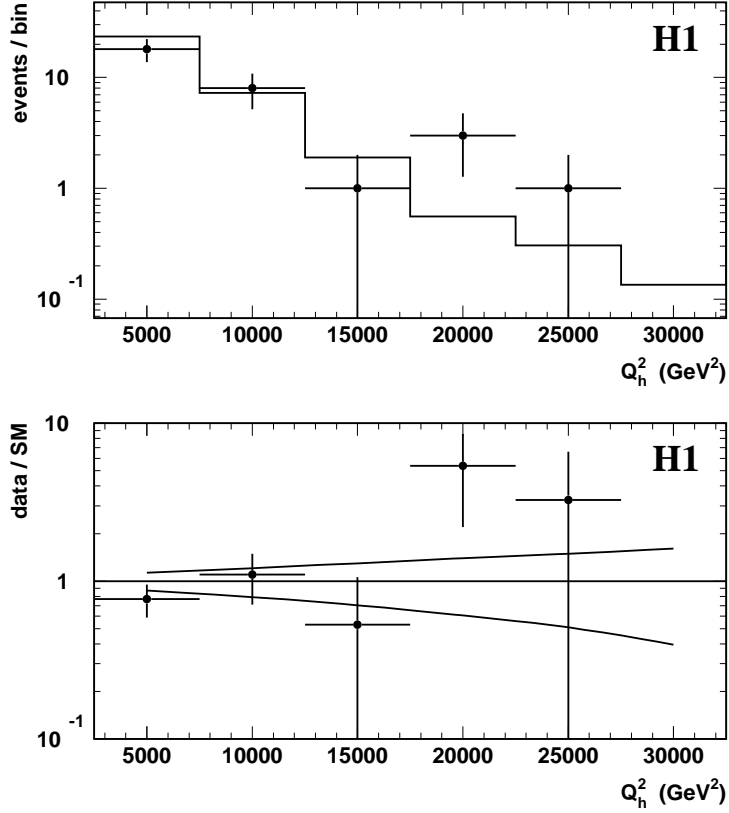}

   \caption[]{ \label{fig:ccq2plot}
      {\small (a) $Q^2_h$ distribution of the selected CC DIS candidate 
              events for the data ($\bullet$ symbols) and for 
              standard CC DIS expectation (histogram); 
              (b) ratio of the observed and expected number of
              events as a function of $Q^2_h$;
              the lines above and below unity specify the $\pm 1\sigma$ 
              levels determined using the combination of statistical and 
              systematic errors of the CC DIS expectation. }}     
\end{figure}

Fig.~\ref{fig:ccq2plot} shows for the CC selection the
measured $Q^2_h$ distribution in comparison with the standard DIS CC 
expectation.
% ---------------------- TABLE 3: Q^2 table for CC  ----------------------------
\begin{table*}[htb]
  \renewcommand{\doublerulesep}{0.4pt}
  \renewcommand{\arraystretch}{1.2}
 \vspace{-0.5cm}

 \begin{center}
 \begin{tabular}{||l|c|c|c|c|c||}
 \hline \hline
$Q^2_{min} (\GeV^2)$  & 2500   & 5000   & 10000  & 15000  & 20000 \\
                                                            \hline \hline
$N_{obs}$             & 31     & 24     & 10     & 4      & 3     \\
                                                            \hline
$N_{DIS}$             & 34.2   & 21.1   & 5.07   & 1.77   & 0.74    \\
    & $\pm 5.8$  & $\pm 4.2$ & $\pm 1.88$  & $\pm 0.87$ & $ \pm 0.39$ \\       
\hline
${\cal{P}}(N \geq N_{obs})$
                      &  0.64  & 0.31   & $7 \times 10^{-2}$  &  0.14   
                                              & $5.4 \times 10^{-2}$  \\
 \hline \hline
  \end{tabular}
         \caption {\small \label{tab:tabq2cc}
                   Number of observed ($N_{obs}$) and expected CC DIS
                   events ($N_{DIS}$) with $4$-momentum 
                   transfer squared $Q^2$ above given $Q^2_{min}$ values; 
                   ${\cal{P}}(N \geq N_{obs})$ is the probability that the 
                   number of CC DIS events fluctuates to values 
                   equal to or larger than $N_{obs}$ in a random set of
                   experiments.}
 \end{center}
\end{table*}
% ------------------------------------------------------------------------
The systematic errors are relatively large and dominated by the uncertainty 
on the hadronic energy scale of the calorimeter. 
% --------------- FIGURE 6: Q2 plot for CC selection ---------------------
%
%-----------------------------------------------------------------------
They vary between $12\%$ at $Q^2_h \simeq 5000 \GeV^2$ 
and $60\%$ at $Q^2_h \simeq 30000 \GeV^2$.
Within errors, the distribution of the measured events is reproduced 
both in shape and in absolute normalization. 
Table~\ref{tab:tabq2cc} gives the number of observed and expected CC events, 
as well as the Poisson probability ${\cal P}(N\geq N_{obs})$ as defined 
in section~\ref{sec:NCsample}, for increasing values of $Q^2_{min}$. 
In the kinematic region $Q^2_h>15000 \GeV^2$, there are $N_{obs}=4$ 
observed events compared with an expectation of $1.77 \pm 0.87$ 
from standard CC DIS. 
This small difference corresponds to a Poisson probability 
${\cal P}(N \geq N_{obs})$ of $14\%$, including systematic errors, 
that the CC DIS signal fluctuates to values equal to  
or larger than $N_{obs}$ in a random set of experiments.

%===================================================================
\subsection{Mass Dependence as a Function of {\boldmath $y$}}

\begin{figure}[tb]

\vspace{-0.5cm}

  \hspace{2.0cm} \epsfxsize=0.7\textwidth \epsffile{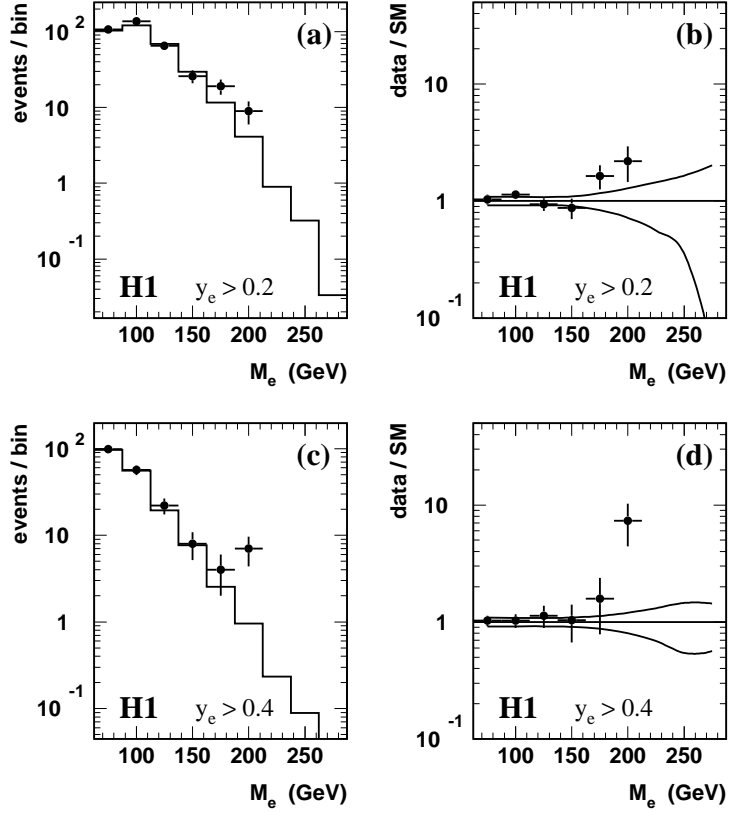}

 \caption[]{ \label{fig:massy0204}
    {\small (a) Distribution of $M_e$ for the observed ($\bullet$ symbols)
            NC DIS candidates with $y_e > 0.2$;
            the expectation from standard NC DIS is shown as the 
            superimposed histogram;
            (b) ratio of the observed and expected numbers of 
            events as a function of $M _e$ for $y_e > 0.2$;
            the lines above and below unity specify the 
            $\pm 1\sigma$ levels of uncertainty
            for the standard NC DIS expectation using the combination of 
            statistical and systematic errors;
            (c) as in (a) but now with $y_e>0.4$;
            (d) as in (b) but now with $y_e>0.4$. }}
\end{figure}

Fig.~\ref{fig:massy0204}a shows the measured and expected $M_e$ 
distribution and Fig.~\ref{fig:massy0204}b the ratio of the measured 
$M_e$ distribution to NC DIS expectation for a minimum $y_e$ value of
$y_{min} = 0.2$. 
Similar distributions are shown in Fig.~\ref{fig:massy0204}c 
and~\ref{fig:massy0204}d for $y_{min} = 0.4$.
An excess of events over the NC DIS expectation at the highest mass 
($\sim 200 \GeV$) is seen, which becomes more visible with the larger
$y_{min}$ cut.
%
% ---------------- FIGURE 7: Mass plot y > 0.2, 0.4 --------------------
%
% -----------------------------------------------------------------------

To quantify the difference in the $M_e$ distribution between the data and
the expectation, the number of events with $M_e$ and $y_e$ above 
various sets of minimal values $M_{min}$ and $y_{min}$ are given in 
Table~\ref{tab:tabmvsy}.
%
% ------------------------ TABLE 4: M vs y -------------------------------
%
\begin{table*}[ht]
\vspace{-0.5cm}

  \renewcommand{\doublerulesep}{0.4pt}
  \renewcommand{\arraystretch}{1.0}
 \begin{center}

 \begin{tabular}{||r|c|c|c|c|}
 \hline \hline

$M_{min}$ (GeV)  $\backslash$  $y_{min}$       
                       & 0.2   & 0.3    & 0.4    & 0.5 \\
                                                         \hline \hline
\begin{tabular}{lr} 
   75 \,\, & $N_{obs}$  \\ & $N_{DIS}$ \\ & ${\cal P}$ \\
\end{tabular} 
 & 
\begin{tabular}{c}
 $ 313 $     \\   $ 294.1 \pm 26.2 $   \\  0.27  \\
\end{tabular} 
 & 
\begin{tabular}{c}
 $ 232 $     \\   $ 208.1 \pm 16.9 $   \\  0.14 \\ 
\end{tabular} 
 & 
\begin{tabular}{c}
 $ 147 $     \\   $ 136.9 \pm 11.2 $   \\  0.26 \\ 
\end{tabular}  
 & 
\begin{tabular}{c}
 $  84 $     \\   $ 77.1 \pm 5.9 $     \\  0.25 \\
\end{tabular} \\
\hline \hline
\begin{tabular}{lr}
 150 \,\, &  $N_{obs}$  \\ & $N_{DIS}$  \\  & ${\cal P}$ \\
\end{tabular}
 & 
\begin{tabular}{c}  
 $  34 $    \\  $ 28.0 \pm 4.6 $     \\  0.19 \\
\end{tabular}
 & 
\begin{tabular}{c} 
 $  17 $    \\  $ 12.9 \pm 1.6 $   \\  0.18 \\
\end{tabular}
 & 
\begin{tabular}{c}
 $ 12  $    \\  $  6.83 \pm 0.80 $    \\ $ 5.3 \times 10^{-2}$ \\
\end{tabular}
 & 
\begin{tabular}{c} 
 $  8  $    \\  $ 3.51 \pm 0.40$  \\ $ 3.0 \times 10^{-2}$ \\
\end{tabular} \\
\hline \hline
\begin{tabular}{lr}
 175 \,\, &  $N_{obs}$  \\ & $N_{DIS}$  \\  & ${\cal P}$ \\
\end{tabular}
 & 
\begin{tabular}{c}  
 $  14 $    \\  $ 9.82 \pm 2.62 $   \\  0.18 \\
\end{tabular}
 & 
\begin{tabular}{c} 
 $  9 $    \\  $ 4.65 \pm 0.77 $   \\  $5.8 \times 10^{-2}$ \\
\end{tabular}
 & 
\begin{tabular}{c}
 $  7 $   \\  $ 2.35 \pm 0.38 $  \\ $ 1.3 \times 10^{-2}$ \\
\end{tabular}
 & 
\begin{tabular}{c} 
 $  5  $    \\  $ 1.31 \pm 0.21 $  \\ $ 1.2 \times 10^{-2}$ \\
\end{tabular} \\
\hline \hline
\begin{tabular}{lr}
 180 \,\, & $N_{obs}$  \\ & $N_{DIS}$  \\ & ${\cal P}$ \\
\end{tabular}
 & 
\begin{tabular}{c}
  $ 11  $      \\ $ 7.85 \pm 2.35 $      \\ 0.22 \\
\end{tabular}
  & \begin{tabular}{c}
 $ 8  $     \\   $ 3.73 \pm 0.70 $    \\  $ 4.6 \times 10^{-2} $ \\
 \end{tabular}
& \begin{tabular}{c}
 $ 7   $     \\ $ 1.83 \pm 0.33 $     \\ $ 3.8 \times 10^{-3}$ \\
\end{tabular}
 & \begin{tabular}{c} 
 $ 5   $     \\ $ 1.04 \pm 0.17$     \\ $ 5.0 \times 10^{-3}$ \\
 \end{tabular} \\
\hline \hline
\begin{tabular}{lr}
 185 \,\, & $N_{obs}$  \\ & $N_{DIS}$  \\ & ${\cal P}$ \\
\end{tabular}
 & 
\begin{tabular}{c}
 $ 10  $      \\ $ 6.10 \pm 2.06 $      \\ 0.14 \\
\end{tabular}
  & \begin{tabular}{c}
 $ 7 $     \\   $ 2.89 \pm 0.57 $    \\  $ 3.6 \times 10^{-2} $ \\
 \end{tabular}
& \begin{tabular}{c}
 $ 7 $     \\ $ 1.41 \pm 0.26 $      \\ $ 9.4 \times 10^{-4} $  \\
\end{tabular}
 & \begin{tabular}{c} 
 $ 5   $     \\ $ 0.83 \pm 0.15 $     \\ $ 2.0 \times 10^{-3}$ \\
 \end{tabular} \\
\hline \hline
\begin{tabular}{lr}
 190 \,\, & $N_{obs}$  \\ & $N_{DIS}$  \\ & ${\cal P}$ \\
\end{tabular}
 & 
\begin{tabular}{c}
 $ 8  $     \\ $ 4.65 \pm 1.68 $      \\  0.15 \\
\end{tabular}
 & 
\begin{tabular}{c}
 $ 6  $    \\ $ 2.19 \pm 0.48 $    \\ $ 3.1 \times 10^{-2} $ \\
\end{tabular}
 & 
\begin{tabular}{c}
 $ 6  $    \\ $ 1.07 \pm 0.23 $     \\ $ 1.2 \times 10^{-3}$ \\
\end{tabular}
 & 
\begin{tabular}{c}
 $ 4  $     \\ $ 0.68 \pm 0.13$     \\ $ 5.9 \times 10^{-3}$ \\
\end{tabular} \\
  \hline \hline
\begin{tabular}{lr}
 195 \,\, & $N_{obs}$  \\ & $N_{DIS}$  \\ & ${\cal P}$ \\
\end{tabular}
 & 
\begin{tabular}{c}
 $ 6  $     \\ $ 3.58 \pm 1.43 $      \\  0.19 \\
\end{tabular}
 & 
\begin{tabular}{c}
 $ 5  $    \\ $ 1.73 \pm 0.40 $    \\ $ 3.8 \times 10^{-2} $ \\
\end{tabular}
 & 
\begin{tabular}{c}
 $ 5  $    \\ $ 0.85 \pm 0.19 $     \\ $ 2.4 \times 10^{-3}$ \\
\end{tabular}
 & 
\begin{tabular}{c}
 $ 4  $     \\ $ 0.53 \pm 0.11$     \\ $ 2.5 \times 10^{-3}$ \\
\end{tabular} \\
  \hline \hline
\begin{tabular}{lr}
 200 \,\, & $N_{obs}$  \\ & $N_{DIS}$  \\  & ${\cal P}$ \\
\end{tabular}
& 
\begin{tabular}{c}
  $ 4   $       \\ $ 2.72 \pm 1.10 $      \\  0.30 \\
\end{tabular}
 & \begin{tabular}{c}
 $ 3  $         \\ $ 1.32 \pm 0.33 $    \\ 0.15 \\
 \end{tabular}
 & \begin{tabular}{c}
  $ 3   $       \\ $ 0.63 \pm 0.15 $     \\ $ 2.8 \times 10^{-2}$ \\
 \end{tabular}
 & \begin{tabular}{c}
   $ 3  $       \\   $ 0.39 \pm 0.09 $     \\  $ 8.2 \times 10^{-3}$ \\
 \end{tabular} \\
 \hline \hline
  \end{tabular}
         \caption {\small \label{tab:tabmvsy}
                   Numbers of observed ($N_{obs}$) NC DIS candidates and 
                   expected ($N_{DIS}$) standard NC DIS events which satisfy 
                   the NC DIS selection for different minimal requirements
                   $ y_e \geq y_{min} $ and $ M_e \geq M_{min} $;
                   also given is the probability 
                   ${\cal{P}}={\cal{P}}(N \geq N_{obs})$ 
                   that the number of DIS events fluctuates to values 
                   equal to or larger than $N_{obs}$ in a random set of
                   experiments.}
 \end{center}
\end{table*}
% ------------------------------------------------------------------------
Also given are the Poisson probabilities ${\cal P}(N \geq N_{obs})$
that the standard NC DIS signal $N$ fluctuates to values equal to or larger 
than the number of observed events in a random set of experiments.
The errors on the NC expectation as well as the probabilities take into
account all systematic errors described in section~\ref{sec:dismc}.
It is noted that the values obtained for the various sets of $M_{min}$ 
and $y_{min}$ cuts are correlated.  
Good agreement is observed for low $M_{min}$ and $y_{min}$, 
whereas at high $M_{min}$ and $y_{min}$ the probabilities 
${\cal P}(N \geq N_{obs})$ become smaller. 

\begin{figure}[t]
\vspace{-1.5cm}

%  \begin{center}
%   \epsfxsize=0.9\textwidth \epsffile{fig.v02.proba.eps}
%   \epsfxsize=0.9\textwidth \epsffile{fig.v02.probab.eps}
   \hspace{1.0cm} \epsfxsize=0.83\textwidth \epsffile{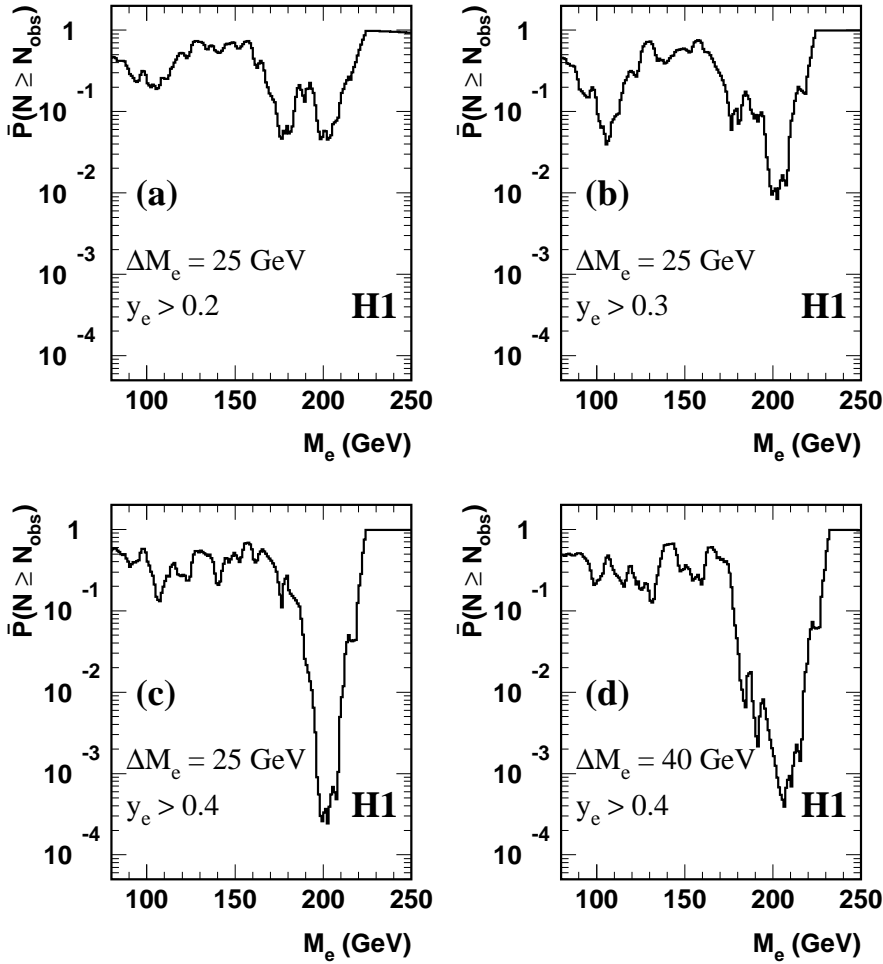}
%  \end{center}

%
 \caption[]{ \label{fig:proba}
    {\small  Probability  that the standard NC DIS signal fluctuates to 
             values equal to or larger than the number of observed events 
             in a random set of experiments for various $y_{min}$ values. 
             The probability is calculated for different sliding mass 
             windows of widths $\Delta M_e$ and averaged over 3 steps of 
             $1 \GeV$.               
             At the highest masses, where no data events are observed, 
             the probabilities approach unity by construction. }} 
\vspace{-0.2cm}
\end{figure}

The range of $M_e$ values for which the most significant excess over NC DIS 
expectation exists is investigated in detail by considering ``windows''
of various total widths $\Delta M_e$.
The central values $M_e$ of the mass windows are varied in steps of 
$1 \GeV$ between $80$ and $250 \GeV$ and the numbers of observed and 
expected events are determined for different $y_{min}$ values.
For each mass window, the Poisson probability ${\cal P}(N \geq N_{obs})$
is determined, including the propagation of all systematic errors described 
above.
The probability for each value of $M_e$ reflects the level of agreement 
between data and expectation for an {\it a priori} choice of $M_e$.
To avoid the inevitable discontinuities when values of ${\cal P}$ are 
displayed due to the small number of observed events, 
a probability ${\bar {\cal P}}$ is calculated averaging over $3$ steps 
of $1 \GeV$ around the given central $M_e$ value.
The resulting Poisson probabilities as a function
of the central value $M_e$ are shown in Fig.~\ref{fig:proba} for
a representative choice of $\Delta M_e$ and $y_{min}$ settings.
This $3 \GeV$ range corresponds to the $M_e$ spread
due to the finite resolution on the measurement of the scattered 
positron energy (see Fig.~\ref{fig:massmeth}d).

For low $y_{min}$, the probabilities fluctuate statistically as expected. 
At masses above $\sim 220\GeV$ no data event is observed within the 
mass windows, and the probabilities ${\bar {\cal P}}(N\geq N_{obs})$ 
approach unity by construction. 
There remain $\sim 0.3$ NC DIS events expected in this high mass region 
for $y_{min}=0.4$.
% --- FIGURE 8: Probability estimates -----------------------------------
%
%-----------------------------------------------------------------------
For $y_{min}$ values of $0.3$ and $0.4$ and for a window width 
varying between $25$ and $40 \GeV$, the lowest probabilities 
of  $10^{-2}$ to $2.6 \times 10^{-4}$
are observed for a central mass value $M_e \sim 200 \GeV$.
For $y_{min} \sim 0.4$, the probabilities are below $8 \times 10^{-4}$ 
for $\Delta M_e = 25\GeV$ and central $M_e$ values chosen {\it a priori} in
the range $\sim 198 \GeV$ to $\sim 208 \GeV$. 
This statement holds with or without averaging ${\cal P}$ over up to 
$5$ steps of $1 \GeV$.
It should be noted that the systematic error on the energy calibration, 
which implies an uncertainty on the number of expected NC DIS events 
and is taken into account in the statistical calculations, gives rise 
to an uncertainty of approximatively $\pm 5\GeV$ for the position 
of the minima in Fig.~\ref{fig:proba}.
The results obtained for a fixed central mass value of $200 \GeV$ with
$y_{min} = 0.4$ and for various choices of $\Delta M_e$, are given in
Table~\ref{tab:tabsigni}.
% ---------------------- TABLE 5: Mass Bin Width -------------------------
\begin{table*}[htb]
  \renewcommand{\doublerulesep}{0.4pt}
  \renewcommand{\arraystretch}{1.2}
 \begin{center}
 \begin{tabular}{||l|c|c|c|c||}
 \hline \hline
$\Delta M_e$ ($\GeV$)
  & $20$         & $25$           &   $30$       & $40$ \\
                                                            \hline \hline
 $N_{obs}$
  & 5            &  7             &  7           &  7           \\
                                                            \hline
 $N_{DIS}$
  & $0.63$      & $0.95$          & $1.10$       & $1.57$       \\
  & $\pm 0.13$  & $\pm 0.18$      & $\pm 0.19$   & $\pm 0.28$   \\
    \hline
${\bar {\cal P}}(N \geq N_{obs})$
           & $5.0 \times 10^{-4}$
                        & $ 2.6 \times 10^{-4}$
                                         & $ 2.5 \times 10^{-4}$
                                                   & $1.6 \times 10^{-3}$ \\
\hline \hline
  \end{tabular}         
  \caption {\small \label{tab:tabsigni}
                   Number of observed NC DIS candidates ($N_{obs}$) and
                   expected standard NC DIS events ($N_{DIS}$) 
                   for different mass windows of total width $\Delta M_e$,
                   an {\it a priori} choice for the central mass value of 
                   $200 \GeV$ and for $y_e > 0.4$; also given are the
                   probabilities ${\bar {\cal P}}(N \geq N_{obs})$ 
                   as defined for Fig.~\protect\ref{fig:proba}.} 
 \end{center}
\end{table*}
% ------------------------------------------------------------------------

In order to estimate the likeliness that in a random experiment a value of 
${\bar {\cal P}}$ smaller than the observed probabilities is obtained 
anywhere in the mass range from $80$ to $250 \GeV$, a large number of 
Monte Carlo experiments were performed.
For each of these experiments, events were randomly chosen according to 
the NC DIS expectation.  
The mean number of events in these experiments was conservatively 
taken to be equal to the number of observed events in the 
corresponding $M_e$ and $y_e$ range.
Applying the same sliding mass procedure as used above for the real 
experiment and comparing with NC DIS expectation including error
propagation, less than $1\%$ of all Monte Carlo experiments yielded a 
minimum value ${\bar {\cal P}}$ below those obtained in 
Fig.~\ref{fig:proba}c and \ref{fig:proba}d.

%===============================================
\subsection{The Events at Very High {\boldmath $Q^2$}, {\boldmath $M$}
            and {\boldmath $y$}}
\label{sec:extreme}

We find $7$ events in the kinematic region $M > 180 \GeV$ and 
$y > 0.4$ compared with an expectation from NC DIS of 
$1.83 \pm 0.33$ ($1.75 \pm 0.32$) using the $e$-method ($2\alpha$-method) 
%-------------- FIGURE 1: Nice ep -> e + X candidate  --------------------
\begin{figure}[t] 

\vspace{-1.5cm}

\hspace*{1.2cm} \epsfxsize=1.0\textwidth \epsffile{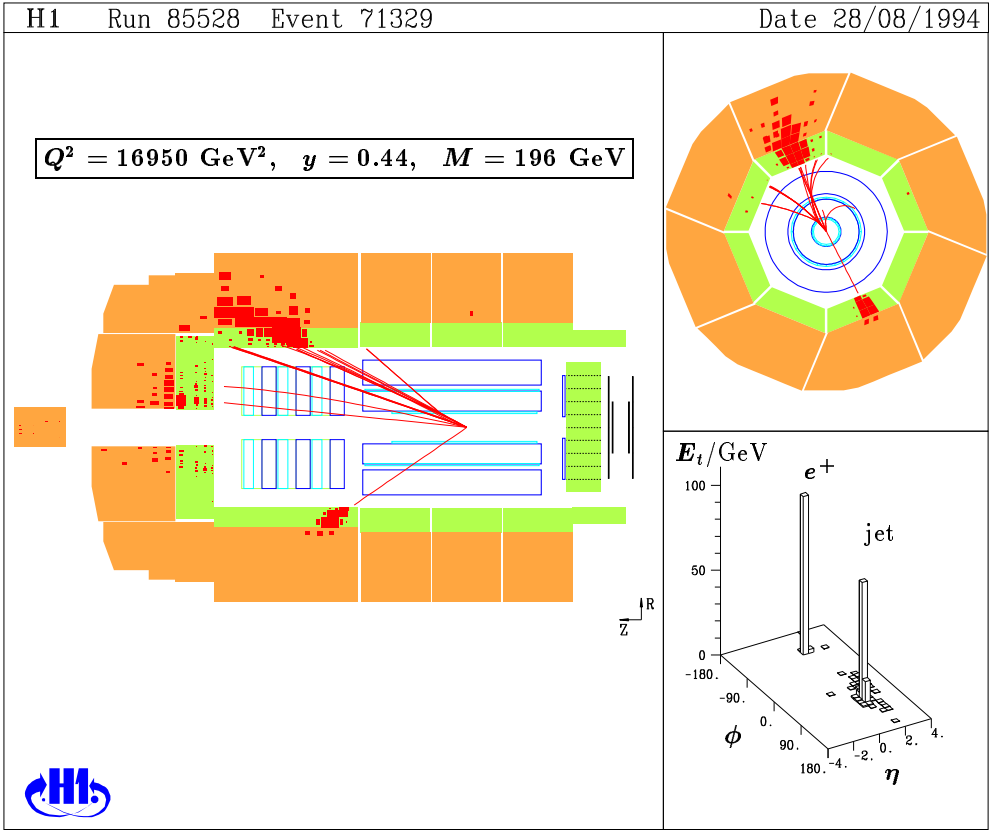}

  \vspace{0.0cm} 

   \caption[]{ \label{fig:event}
      {\small  An event of the NC DIS sample at 
               $M_e > 180 \GeV$ and $y_e > 0.4$. }} 
\end{figure}
%-------------------------------------------------------------------------
These events are largely responsible for the excess of observed
events over standard DIS expectation in the $Q^2$, $M$ and $y$ distributions 
in Fig.~\ref{fig:mandy},~\ref{fig:q2plot} and~\ref{fig:massy0204}.
They are listed in Table~\ref{tab:events} and their properties are 
discussed here in more detail.
One of these events is displayed in Fig.~\ref{fig:event}.
%
% ---------------------- TABLE 2: Event List ------------------------------
%
\begin{table*}[htb]
  \renewcommand{\doublerulesep}{0.4pt}
  \renewcommand{\arraystretch}{1.2}
 \begin{center}
 \begin{tabular}{||c|l|c|c|c|c|c|c|c||}
 \hline \hline
    &  Run \#      & $M_e$ &  $M_{2\alpha}$  &  $M_{e+jet}$  
                                             & $y_e$   & $y_{2\alpha}$ 
                                             & $Q^2_e$ & $ Q^2_{2\alpha}$ \\
    &  Event \#    & (GeV) &    (GeV)        &    (GeV) 
                                             &         & 
                                             & ($\GeV^2$) &($\GeV^2$)    \\ 
 \hline
 1  & 85528  & 196     & 198 & 195 & .439       & .434 & 16950     & 17100 \\
    & 71329  & $\pm$5  &     &     & $\pm$.014  &      & $\pm$360  &       \\
 2  & 87050  & 208     & 200 & 213 & .563       & .582 & 24350     & 23320 \\
    &  8062  & $\pm$4  &     &     & $\pm$.012  &      & $\pm$430  &       \\
 3  &  88999 & 188     & 185 & 184 & .566       & .573 & 19950     & 19640 \\
    & 106218 & $\pm$12 &     &     & $\pm$.032  &      & $\pm$1400 &       \\
 4  & 119314 & 198     & 199 & 197 & .790       & .787 & 30870     & 31320 \\
    &  18272 & $\pm$2  &     &     & $\pm$.008  &      & $\pm$530  &       \\
 5  & 122145 & 211     & 227 & 211 & .562       & .526 & 25030     & 27100 \\
    &  69506 & $\pm$4  &     &     & $\pm$.012  &      & $\pm$440  &       \\
 6  & 155768 & 192     & 190 & 215 & .440       & .443 & 16130     & 16050 \\
    &  67216 & $\pm$6  &     &     & $\pm$.016  &      & $\pm$400  &       \\
 7  & 158577 & 200     & 213 & 210 & .783       & .762 & 31420     & 34450 \\
    & 100110 & $\pm$2  &     &     & $\pm$.008  &      & $\pm$540  &       \\
  \hline \hline
  \end{tabular}
         \caption {\small \label{tab:events}
                   NC DIS candidates satisfying the kinematic requirement 
                   $M > 180 \GeV$ and $y > 0.4$ when $M$ and $y$ are 
                   calculated with either the $e$-method or the 
                   $2\alpha$-method;
                   the errors given for the $e$-method take into account 
                   the energy and angular resolution for the positron, and 
                   a $30\%$ uncertainty assigned to dead material 
                   corrections;
                   the systematic uncertainty of
                   $3\%$ associated with the absolute energy scale of 
                   the electromagnetic calorimeter (see text) is not
                   included.}
 \end{center}
\end{table*}
%
% ------------------------------------------------------------------------
%

% Technical Studies

For most of these events, the positron is well contained within the 
fiducial volume of the LAr calorimeter.
The corrections for material in front of or in between calorimeter 
modules, which are applied during the reconstruction of the energy from 
showers in the calorimeter, were found to be less than $1\%$ ($9\%$) except 
for the electromagnetic (hadronic) showers of events 
$3$ and $6$ ($6$).
The electromagnetic showers initiated by the positron of events $3$ and
$6$ are corrected by $23\%$ and $5\%$ respectively because they develop near 
a projective azimuthal edge of a calorimeter module.
The corrections to the energy of the hadronic showers of event
$6$ is $29\%$ also largely due to shower leakage into dead 
material in between calorimeter modules.
The agreement between the different estimators of the kinematic 
variables in Table~\ref{tab:events} demonstrates that the corrections 
are well understood on average. 
In particular, the two estimators $M_e$ and $M_{2\alpha}$ are seen to agree
well with each other and also with the invariant mass $M_{e+jet}$ 
calculated from the positron and the highest $E_T$ jet in the final 
state.

In each event, the identified positron cluster is geometrically linked
to at least one high momentum charged track.
In one case (event $4$), multiple tracks or track segments are 
associated spatially with the positron cluster as expected from 
Monte Carlo studies in $15 \pm 5 \%$ of events with similar kinematics,
given the inactive material between the calorimeter and the $ep$ 
interaction vertex.
The $7$ events are not confined to any particular azimuthal region of
the detector.
One event ($5$) is found to have $2.9 \GeV$ deposited in the photon
detector used to determine luminosity from Bethe-Heitler interactions.  
The probability that one event out of seven of the high $Q^2$, high $y$ 
sample be in random coincidence with an elastic Bethe-Heitler interaction 
with one photon detected is estimated to be $\sim 35\%$. 
None of the $7$ events have high momentum muons which penetrate the
calorimeter and instrumented iron filter. 
For each of the $7$ events, only one jet is found with 
$E_{T,jet} > 15 \GeV$ with the cone algorithm and this jet carries 
more than $90\%$ of the transverse hadronic energy flow.

All the above indicate that these $7$ events are well measured 
NC DIS-like candidates.
From Table~\ref{tab:events}, the weighted average mass $<M_e>$
measured for these candidates is $ 200.8 \pm 2.2 \GeV$.
This value does not include the systematic uncertainty 
of $3\%$  associated with the absolute energy scale of the
electromagnetic calorimeter.
This uncertainty leads to a correlated error for all events which
depends on $y_e$ like $\delta M_e/M_e =0.03/(2y_e)$, 
$\delta y_e = 0.03 (y_e-1)$ and $\delta Q^2_e/ Q^2_e = 0.03$.
 
For the CC DIS sample, in the kinematic region $Q^2_h>15000 \GeV^2$, 
which includes the region in which an excess is observed in NC DIS, 
there are $N_{obs}=4$ events in agreement with the expectation of 
$1.77 \pm 0.87$ from standard CC DIS. 
These $4$ events are listed in Table~\ref{tab:ccevents}.
%
% ---------------------- TABLE: CC Event List ------------------------------
%
\begin{table*}[htb]
  \renewcommand{\doublerulesep}{0.4pt}
  \renewcommand{\arraystretch}{1.2}
 \begin{center}
 \begin{tabular}{||c|c|c|c|c||}
 \hline \hline
%
% ( $Q^2 \gtrsim 15000 \GeV^2$ )
% 

    &  Run \# / Event \#  & $M_h$ & $y_h$ & $Q^2_h$  \\
 \hline
 1  &  85987/ 99058  & 213  & .52  & 23380 \\  
 2  &  153720/199055 & 187  & .58  & 20170 \\  
 3  &  163852/20613  & 157  & .75  & 18650 \\  
 4  &  169851/206239 & 192  & .56  & 20800 \\
  \hline \hline
  \end{tabular}
         \caption {\small \label{tab:ccevents}
                   CC DIS candidates satisfying the kinematic requirement 
                   $Q^2_h > 15000 \GeV^2$;
                   in the $M$, $y$ and $Q^2$ range spanned by these events,
                   the expected resolutions are 
                   $\delta M_h/M_h \sim 10\%$,
                   $\delta y_h/y_h \sim 10\%$,
                   $\delta Q^2_h/Q^2_h \sim 28 \%$, 
                   not including the systematic uncertainty of
                   $4\%$ associated with the absolute hadronic energy 
                   scale.}
 \end{center}
\end{table*}
% ------------------------------------------------------------------------

%=========================================================================
\section{Non-Standard Contributions at High {\boldmath$Q^2$}}
\label{sec:anomalous}

Having presented the essential experimental evidence, we now discuss 
possible contributions to an excess of events at high $Q^2$ 
(high $M$ and $y$) within and beyond the Standard Model.

\subsection{New Contributions in the Standard DIS Model}

The standard DIS expectations given in section~\ref{sec:dismc} take into
account a systematic error due to a specific choice of the parton 
density parametrizations. 
No parton density parametrization compatible with existing experimental 
DIS data was found to lead to an enhancement of the NC DIS expectation at 
high masses ($M \gtrsim 180 \GeV$) for $Q^2 > 15000 \GeV^2$ beyond the 
systematic uncertainty quoted in section~\ref{sec:dismc}.
This is in particular the case for the parametrization 
CTEQ~(4HJ)~\cite{CTEQ4MHJ} which was designed to cope with the apparent
rise of the di-jet cross-section observed at the Tevatron~\cite{TEVAJJ}
and leads to an enhancement, compared to MRS~(H), of $\lesssim 5 \%$ 
in the above cited $M$ and $Q^2$ range.
In the same context, the parametrization MRS~(R2)~\cite{MRSR12} 
which assumes a ``large'' $\alpha_s$ value 
($\alpha_s (M_Z^2) \simeq 0.120$) leads to an enhancement of the scaling 
violation and consequently to a slight reduction of the differential 
cross-section at high $M$ relative to MRS~(H), also within 
systematic errors.

The MRS~(J') parametrization~\cite{MRSJP} attempts to reproduce
the magnitude of the rise of the di-jet cross-section observed in 
particular in CDF~\cite{TEVAJJ} but does not fit existing DIS data
(e.g. BCDMS data at high $x$ and low $Q^2$). It would lead to a cross-section 
about $20\%$ larger than the values used for this analysis which would 
still not explain the excess of events observed in the above cited 
$M$ and $Q^2$ range.

% OTHER QCD MODEL

The statistical significance presented in previous sections for an
excess at high $Q^2$ and/or high $M$ and $y$ is unchanged when using 
the LEPTO~\cite{LEPTO} model for the QCD corrections.

% HIGHER ORDER QED

The expectations from the standard DIS model have been given under
the conservative assumption that the longitudinal structure function
$F_L$ is equal to zero.
A finite value of $F_L$ either originating from QCD or 
due to Fermi motion in the proton, would decrease the cross-section 
at large values of $Q^2$ and $y$. 

% SM CONCLUSION
 
From the above, and given the agreement of the measurement with
NC DIS expectation at low $M$, $Q^2$ and $y$ (see Fig.~\ref{fig:mandy}),
there appears to be no mechanism within the standard DIS model 
framework which would lead to an enhancement of the cross-section 
at high $Q^2$ or high $y$ as observed.
Within the standard DIS model the only explanation of our result is 
therefore a statistical fluctuation.

\subsection{Physics Beyond the Standard Model}

Beyond the Standard Model, the excess at high $Q^2$ and/or high $M$ and $y$
could be explained by different mechanisms.

New particles could be produced as resonances in the positron-parton 
system. Prominent examples for new particles with couplings to positron-parton 
pairs are leptoquarks~\cite{LEPTOQUARK}, leptogluons~\cite{LEPTOGLUON} and 
squarks in R-parity violating versions of Supersymmetry \cite{SQUARK}.
It should be emphasized that the reconstruction of the invariant mass of
such a resonance is hampered by QED  and QCD corrections.
An example of such an effect in the form of a narrow $s$-channel resonance 
is shown in Fig.~\ref{fig:massmeth}d as a dotted histogram. 
The LEGO~\cite{LEGO} generator, which was used to produce such a resonance 
at fixed mass $M_{gen}$, incorporates initial state QED bremsstrahlung in the
collinear approximation and QCD initial and final state parton showers and
fragmentation~\cite{JETSET74}. It moreover properly takes into account the 
effects of the parton shower masses on the decay kinematics.
It can be seen that a resonance would be reconstructed with a systematic
shift (typically $\lesssim 5\%$) towards smaller mass values and with a 
resolution considerably worse than expected from detector effects alone.
At the present level of significance, neither the RMS width of the
measured events of  $8.5 \GeV$ nor the $y_e$ distribution
(Fig.~\ref{fig:mandy}f), which is expected to extend to larger $y_e$
for a resonance than for NC DIS processes, 
allow conclusions to be drawn concerning the possibility of 
resonance formation.

A slight excess is seen in the $Q^2$  distribution 
(Fig.~\ref{fig:q2plot}, Table~\ref{tab:tabq2}).
The virtual exchange of new particles between positrons and partons 
or possible substructure of fermions inducing contact interactions
could all lead to an enhancement of the cross-section at high $Q^2$.

%=======================================================================
\section{Summary}

Deep-inelastic scattering (DIS) events have been observed in $e^+p$ 
collisions at very large $4$-momentum transfer squared $Q^2$, and have 
been compared with the expectations from the standard Neutral Current (NC) 
and Charged Current (CC) DIS model.

%%%> Experimental facts :

For $Q^2 \lesssim  15000 \GeV^2$, the distributions of $Q^2$ or
$M= \sqrt{xs}$ and $y = Q^2/M^2$ are well reproduced by the
expectation of standard DIS.
At larger momentum transfer ($Q^2 > 15000\GeV^2$), $12$ 
NC DIS candidate events are observed where $4.71 \pm 0.76$ are expected 
and $4$ CC DIS candidate events are observed where $1.77 \pm 0.87 $ are 
expected.
The Poisson probability ${\cal {P}}$ that the signal from standard DIS 
fluctuates to a number of events equal to or larger than the observed 
number of events is $6 \times 10^{-3}$ in the NC case and 
$0.14$ in the CC case.

For the NC candidates, the excess of events is most prominent in a mass 
window of total width $25 \GeV$ centered at an invariant mass 
$M \simeq 200 \GeV$ of the positron--parton system.
This mass is consistently determined using different kinematic reconstruction
methods, for which only the measured positron, or only the measured 
angles of the positron and the hadronic system, or only the $4$-momenta of the 
positron and the jet with highest $E_T$ are used.
The dominant $e+{\rm jet}+X$ topological feature of the events is 
characteristic of standard neutral current DIS processes.
In a mass window of $25 \GeV$ width with a central value of $200 \GeV$ and 
for $y>0.4$, 7 events are observed where $0.95\pm0.18$ events are expected. 
For this and other choices of mass windows and $y$ thresholds, 
the probability to observe an excess as large as the measured one anywhere 
in the mass range investigated is of order $1\%$.

%%%%> Physics message :

No known detector effect can account for an excess at large
$Q^2$ or can be associated with an excess which occurs preferentially
in a restricted range of $M$.

Given the existing experimental constraints on parton density 
distributions at high $M$ and lower $Q^2$, and given the agreement
of the resulting predictions for $e^+p$ DIS which is here reported
for $Q^2$ below $15000\GeV^2$,
there is little freedom for an enhancement of the cross-section 
at higher $Q^2$ by different choices for the 
partonic structure of the proton, by changing the strong coupling 
constant, or by including higher order corrections.

A new mechanism would be needed to explain an enhancement of the DIS 
cross-section affecting mostly high $Q^2$ or high $y$ values.  
Within the standard DIS model the only explanation of our result is 
therefore a statistical fluctuation.

Whereas an account of the observation by introducing new physics beyond
the standard model of electroweak and strong forces is kinematically  
possible, the signature of the observed events is identical to DIS 
and hence a clarification of their nature will have to come 
from the study of kinematic distributions with larger statistics.

%=======================================================================
\section*{Acknowledgements}
We wish to thank the HERA machine group as well as the H1 engineers and
technicians who constructed and maintained the detector for their
outstanding efforts.
We thank the funding agencies for their financial support.
We wish to thank the DESY directorate for the support
and hospitality extended to the non-DESY members of the collaboration.
 
%=======================================================================
 
{\Large\normalsize}

\end{document}